\documentclass[aip,amsmath,amssymb,amsfonts,reprint]{revtex4-1}

\usepackage{amsthm}
\usepackage{graphicx} 
\usepackage{MnSymbol}
\usepackage[draft]{changes}
\usepackage[english]{babel}
\usepackage{ucs}
\usepackage[utf8x]{inputenc}
\usepackage{nicematrix}
\usepackage{color,soul}

\newcommand{\mb}[1]{\mbox{\bfseries \itshape #1}}

\newcommand{\N}{\mathbb{N}}
\newcommand{\Z}{\mathbb{Z}}

\newtheorem{defi}{Definition}
\newtheorem*{defiNN}{Definition}

\begin{document}

\title{Templex-based dynamical units for a taxonomy of chaos}

\author{Caterina Mosto}
\email{caterina.mosto@cima.fcen.uba.ar}
\affiliation{CONICET – Universidad de Buenos Aires. Centro de Investigaciones
del Mar y la Atmósfera (CIMA), C1428EGA  CABA, Argentina}
\affiliation{CNRS – IRD – CONICET – UBA. Institut Franco-Argentin d'\'Etudes sur le Climat et ses Impacts (IRL 3351 IFAECI), C1428EGA  CABA, Argentina.}

\author{Gisela D. Char\'o}
\email{gisela.charo@cima.fcen.uba.ar}
\affiliation{CONICET – Universidad de Buenos Aires. Centro de Investigaciones
del Mar y la Atmósfera (CIMA), C1428EGA  CABA, Argentina}
\affiliation{CNRS – IRD – CONICET – UBA. Institut Franco-Argentin d'\'Etudes sur le Climat et ses Impacts (IRL 3351 IFAECI), C1428EGA  CABA, Argentina.}
\affiliation{Laboratoire des Sciences du Climat et de l’Environnement, CEA Saclay l’Orme des Merisiers, UMR 8212 CEA-CNRS-UVSQ, Université Paris-Saclay \& IPSL, 91191, Gif-sur-Yvette, France.}

\author{Christophe Letellier}
\homepage{http://www.atomosyd.net/spip.php?article1}
\email{christophe.letellier@coria.fr}
\affiliation{
Rouen Normandie University --- CORIA, Campus Universitaire du Madrillet,
F-76800 Saint-Etienne du Rouvray, France }

\author{Denisse Sciamarella}
\email{denisse.sciamarella@cnrs.fr}
\affiliation{CNRS – IRD – CONICET – UBA. Institut Franco-Argentin d'\'Etudes sur le Climat et ses Impacts (IRL 3351 IFAECI), C1428EGA  CABA, Argentina.}
\affiliation{CNRS – Centre National de la Recherche Scientifique, 75016 Paris, 
France.}

\date{\today ~to submit to {\it Chaos}}

\begin{abstract}
Discriminating different types of chaos is still a very challenging topic, even for dissipative three-dimensional systems for which the most advanced tool is the template. Nevertheless, getting a template is, by definition, limited to three-dimensional objects, since based on knot theory. To deal with higher-dimensional chaos, we recently introduced the templex combining a flow-oriented {\sc BraMAH} cell complex and a directed graph (a digraph). There is no dimensional limitation in the concept of templex. Here, we show that a templex can be automatically reduced into a ``minimal'' form to provide a comprehensive and synthetic view of the main properties of chaotic attractors. This reduction allows for the development of a taxonomy of chaos in terms of two elementary units: the oscillating unit (O-unit) and the switching unit (S-unit). We apply this approach to various well-known attractors (Rössler, Lorenz, and Burke-Shaw) as well as a non-trivial four-dimensional attractor. A case of toroidal chaos (Deng) is also treated. This work is dedicated to Otto E. Rössler.
\end{abstract}

\maketitle

\begin{quotation}
The nonlinear essence of many (if not most) of the dynamical processes in nature unavoidably led to the discovery of chaos, a deterministic dynamics for which long term prediction is not possible. With the emergence of desktop  computers, many different types of chaos were reported, but a complete taxonomy is still under development. One of the limitations of the topological characterization of chaotic attractors is the use of knot theory, whose  applicability is limited to three-dimensional space. Templexes are objects allowing for a definition of an algebraic topology on a cell complex (which, unlike knots, has no dimensional restrictions) and a directed graph whose nodes  are the locally highest dimensional cells of the complex and whose edges indicate the way in which the flow visits such cells. This work shows that a simplified templex can describe various types of chaotic attractors as  different combinations of two elementary dynamical processes, namely oscillating units and switching units. Such units pop up after the reduction procedure that merges cells or nodes, without altering the topological properties. The way in which the templex is reduced is crucial in order to identify the basic units underlying a given dynamics. Different types of chaotic attractors are discussed.
\end{quotation}

\section{Introduction}

Understanding the structure of chaotic attractors has practical applications in various fields, including physics, biology, economics, and engineering. For instance, chaos theory has been applied in fields such as weather prediction\cite{Lor65,Lor68,She21} and control,\cite{Sun23} fluid dynamics,\cite{Lor63b,Yor85,Are17} optics,\cite{Sun11} secure communication,\cite{Wan02} and the modeling of ecological systems.\cite{Has91,Has93} Many topologically inequivalent chaotic attractors are now known in almost any  field. There is no general relationship between the algebraic structure of the governing equations and the type of attractors they produce. Nevertheless, in the early 1970s, Otto E. Rössler developed complex chemical reactions as the combination of some basic circuits,\cite{Ros72a} an approach  inherited from his background in electronics.\cite{Let10b} He was following the works of Sugita who developed an analogy between chemical systems and logical circuits.\cite{Sug63a,Sug63b} Thus, Rössler introduced his eponym system as a combination between a two-dimensional oscillator and a nonlinear switch.\cite{Ros76a} He described the resulting chaotic attractor as the union of a ``normal'' strip and a Möbius strip, that is, using a branched manifold. The topological approach thus appeared to be the adequate way to describe chaotic attractors and to classify them. He then provided an early attempt to construct a ``hierarchy of chaos'' based on positive Lyapunov exponents.\cite{Ros78,Ros83} This led to the recent taxonomy of chaos\cite{Let22a} by using the number $p$ of positive Lyapunov exponents and the number $q$ of null Lyapunov exponents to construct the marker C$^p$T$^q$ for discriminating different classes of attractors, C$^p$ discriminating periodic or quasi-periodic ($p = 0$) from chaotic ($p = 1$) of hyperchaotic ($p = 2$) behaviors: T$^q$ directly designates the torus T$^q$ on which the structure of the attractor is based (for instance, a limit cycle with $q = 1$ and a torus for $q = 2$).

Topological characterization of chaos provides a deep understanding of the intricate and complex behavior produced by chaotic dynamical systems.\cite{Tuf92,Let95a,Gil98,Gil03} It offers analysis and description of the underlying structure of chaos in a systematic and mathematical way.\cite{Let22a} Topology is a branch of mathematics that deals with the  properties of a spatial object that remains unchanged under isotopy, that is, continuous deformations such as stretching, bending, or twisting. In the context of chaos theory, topological methods are used to analyze and describe the structure of chaotic attractors produced by numerical simulations as well as measured in experiments.\cite{Sol88a,Min92,Lef93,Let95c,Gil97,Let98d,Sci01b} For many years, such a characterization was mainly based on knot theory and the construction of a template as a knot-holder.\cite{Bir83a} A template is a branched manifold sketched with a splitting chart, some strips with some local torsions and permutations. These strips are finally squeezed at the so-called joining chart. The way in which strips are squeezed can follow a ``standard'' convention as introduced by Tufillaro and coworkers.\cite{Tuf92} Characterizing higher-dimensional attractors was therefore a great challenge in spite of a few attempts.\cite{Lef06,Lef13} Indeed, building a template is strongly related to the relative organization of unstable periodic orbits extracted from the chaotic attractors, an organization which is quantified using topological invariants which are only valid in three-dimensional space.

A very promising approach consisted in borrowing concepts from homology group theory, such as Betti numbers,\cite{Mul93} to study a trajectory in the state space or in a reconstructed space. This approach was pushed further by \citet{Sci99,Sci01b} who used cell complexes to describe branched manifolds in terms of homology and torsion groups. A cell complex is a layered set of cells of different dimensions designed to classify and describe the shape of a manifold up to topological equivalence -- see appendix \ref{algebraictop} for a few definitions and examples. This specific type of cell complex is called {\sc BraMAH},\cite{Cha20} an acronym standing for Branched Manifold Analysis through Homologies. A {\sc BraMAH} complex is built in a particular fashion, so that its highest dimensional cells locally approximate subsets of points lying on a branched manifold: it does not rely on unstable periodic orbits (knots) and, consequently, there is no intrinsic limitation. Recent developments of this approach have led to the definition of an unprecedented mathematical object,\cite{Cha22} called templex. 

The key feature is that the complex is endowed with a directed graph, whose nodes are identified with the locally highest dimensional cells of the complex. The directed graph specifies how the flow visits the structure approximated by the complex. Of course, this only has sense for a complex produced by a dynamical system. A templex gathers in a single mathematical object a set of algebraic properties which describes the topology of the cell complex and of the flow upon it. Here, we will extend the potential of templex in showing that a chaotic system results from a combination between oscillators and nonlinear switches obtained after a reduction procedure, which decreases the number of cells in the complex and, consequently, of nodes in the digraph. Indeed, the reduced digraph provides a systematic way to describe chaotic attractors in terms of (at least two-dimensional) oscillators (O-units) and (at least one-dimensional) switches (S-units). 

The subsequent part of this paper is organized as follows. Section \ref{temple} introduces the concepts underlying templex. Section \ref{dynit} describes the O- and S-units and how they are related to a taxonomy of chaos. Section \ref{dynamit} describes how a complex is reduced and the way to extract the units. In Section \ref{example}, a set of very different attractors (the spiral and funnel Rössler attractors, the Lorenz attractor, the Burke and Shaw attractor, a non-trivial 4D attractor and the Deng toroidal chaos) are treated. Section \ref{conc} gives some conclusions.

\section{Theoretical background}

\subsection{Templex}
\label{temple}

A templex is based on a cell complex constructed from a cloud of points corresponding to a trajectory produced by a dynamical system. A {\sc BraMAH} complex is built using this specificity as initiated by one of us,\cite{SciPhD,Sci99,Sci01b} and later improved.\cite{Cha20,Cha21,Cha22} Cell  complexes provide a combinatorial approach to topology.\cite{Kin93} A cell complex $K$ of dimension $\kappa$ ($\kappa \in \mathbb{N}$) is a layered structure built up of a finite set of $k$-cells ($k= 0, 1, \hdots \kappa$): $0$-cells are points, $1$-cells are segments, $2$-cells are filled polygons, $3$-cells are solid polyhedra, and so forth. The dimension of the complex $K$ is given by the dimension of its highest-dimensional cells. In a given attractor, it is possible that, depending on the location in the state space, the highest-dimensional cells is such that $k < \kappa$. It is therefore useful to introduce the local dimension $\kappa_{\rm l} (\mb{x})$ of the highest-dimensional cell at point $\mb{x}$:  consequently, $\kappa = \left. \displaystyle \mbox{max } \kappa_{\rm l} (\mb{x}) \right|_{\mb{x} \in \mathbb{R}^d}$. The topological properties of such a cell complex are independent of the number, shape or distribution of the cells: what matters is how they are glued together. These properties are expressed by its homology groups $H_k(K)$ and its torsion coefficients. The $k$-generators of the homology groups are also called $k$-holes, since they are non-trivial chains of $k$-cells that encircle empty areas ($k=1$), empty volumes ($k=2$) or empty hyper-volumes ($k \ge 3)$. The  $0$-holes denote the number of connected components in the cell complex. In order to make a cell complex as simple as possible, a natural question is whether there is a minimal number of $k$-cells to approximate the attractor underlying the trajectory. 

In a dynamical system, there is a flow underlying the cloud of points used to build the cell complex and therefore, there is a deterministic sequence for visiting the cells. This can be schemed as a directed graph, here designated as a digraph.\cite{Cha22} When the dynamics of the flow is not regular (where regular means periodic or quasi-periodic), the flow is structured by a branched manifold as required for chaotic behavior. A branched manifold is characterized by splitting and joining loci\cite{Cha22} which are necessary to describe the non equivalent paths followed by the flow. The case of regular dynamics is briefly considered in Appendix \ref{regdyn}.

\begin{figure}[ht]
  \centering
  \begin{tabular}{ccc}
    \includegraphics[width=0.18\textwidth]{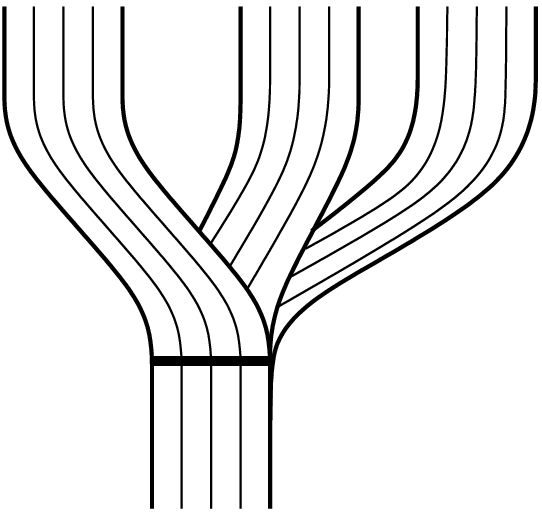} &
    \includegraphics[width=0.20\textwidth]{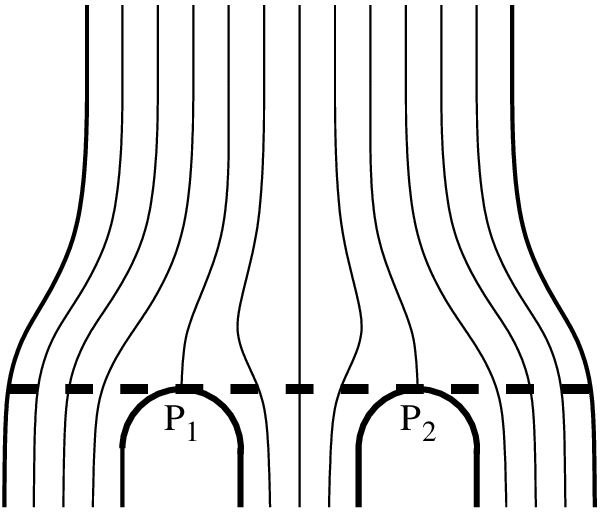} \\
    {\small (a) Joining locus} & {\small (b) Splitting locus} \\[0.2cm]
    \includegraphics[width=0.10\textwidth]{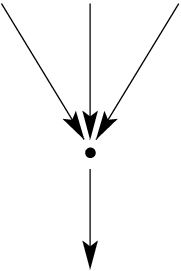} &
    \includegraphics[width=0.10\textwidth]{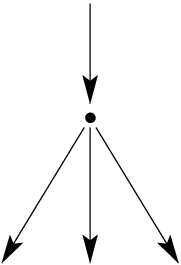} \\
    {\small (c) Joining subgraph} & {\small (d) Splitting subgraph} \\[-0.2cm]
  \end{tabular}
  \caption{Three ingoing strips --- the flow is from the top to the bottom --- are joined into a single outgoing strip in (a). An ingoing strip is split into three outgoing strips according to the two critical points P$_1$ and P$_2$ in (b). In (a), boundaries are in solid thick lines. The thick line in (a) represents the joining locus and the dashed thick line in (b) corresponds to the splitting locus. In (c), joining subgraph. In (d), splitting subgraph. } 
  \label{splitjoin}
\end{figure}

A joining locus exists in a {\sc BraMAH} complex of dimension $\kappa$ when there are at least three $\kappa$-cells sharing a ($\kappa -1$)-cell. When $\kappa = 2$ as for a branched $2$-manifold, the joining locus is a joining line. Let us designate as {\it ingoing} ({\it outgoing}) $\kappa$-cells, the $\kappa$-cells visited by the flow just before (after) crossing one of the joining ($\kappa-1$)-cells belonging to a joining locus. As for strips in a template, a joining locus must have at least two ingoing and a single outgoing cell [Fig.\ \ref{splitjoin}(a,~c)]; a splitting locus allows to have at least two $\kappa$-cells outgoing from a single ingoing $\kappa$-cell, [Fig.~\ref{splitjoin}(b,~d)]. 

To be able to compute homology and torsion groups, cell complexes must be directed or oriented, that is, each $k$-cell ($k>0$) must be assigned an arbitrary direction, regardless of whether there is an underlying flow. These orientations enable defining an arithmetic on the cells and thus an algebra of chains. In order to build a templex, there are special rules regarding how the highest dimensional cells should be oriented around the joining loci. This is performed by orienting the $\kappa$-cells that are outgoing from a joining locus according to the flow; the direction of its peripheral boundary being along the flow. The orientation is then propagated from $\kappa$-cell to $\kappa$-cell with an important caveat: orientation cannot be propagated across a joining locus without ambiguity. It is therefore sometimes necessary to restart the orientation process by repeating the procedure from a not yet oriented cell until completion of the orientation.

Each joining ($\kappa-1$)-cell must be oriented according to the outgoing $\kappa$-cell from which it is a boundary. If there is a change in direction within a connected set of joining $(\kappa-1)$-cells, then there is a {\it critical} $(\kappa-2)$-cell which divides this connected set into two different connected sets of joining $(\kappa-1)$-cells. The existence of such a critical $(\kappa-2)$-cell is induced by our convention for orienting the $\kappa$-cells according to the flow. This is strongly related to the convention for orienting the components of the Poincaré section from the center to the periphery to avoid different descriptions of the same dynamics.\cite{Ros13} Once this is completed, each connected set of $(\kappa-1)$-cells sharing the same direction corresponds to one component of the Poincaré section. Each component is denoted by $J_i$ where $i \in \mathbb{N}$. The concept of multi-component Poincaré section was introduced for systems with symmetry,\cite{Let94a,Let01,LetPhD} and then generalized by Tsankov and Gilmore.\cite{Tsa04} When $i = 2$, there is a correspondence between templex and template in the dissipative three-dimensional limit which 
was discussed in Ref. \onlinecite{Cha22}.

Once the $\kappa$-dimensional complex is built and oriented, the directed graph must encode the information regarding the way $\kappa$-cells are visited by the underlying flow. A digraph is made up of vertices or nodes, which are connected by directed boundaries, also called edges. We will use hereafter the terms nodes and edges. Two $\kappa_{\rm l}$-cells or nodes $i$ and $j$ are connected by a directed edge as $i \rightarrow j$ when the flow switches from a $\kappa_{\rm l}$-cell $\gamma_i$ to a $\kappa_{\rm l}$-cell $\gamma_j$. This naturally leads to the definition of {\it templex}.\cite{Cha22}

\begin{defi}
A {\bf templex} $T = (K,G)$ is made of a {\sc BraMAH} complex $K$ with dim$(K)=\kappa$ and a digraph $G=(N,L)$ whose underlying space is a branched $\kappa$-manifold associated with a dynamical system, such that (i) the nodes $N$ are $\kappa_{\rm l}$-cells and (ii) the edges $L$ are the connections between the $\kappa_{\rm l}$-cells visited by the flow.
\end{defi}

\noindent
The advantage of a templex over a cell complex is that its algebraic descriptors include both the topology of the structure (complex) and of the organization of the flow on that structure (digraph). The topology of the structure is described in terms of homologies and torsions as in traditional topology textbooks, allowing for the distinction between, for instance, a torus and a Klein bottle. The topology of the flow on the cell complex is described through a set of definitions that enable computing the non-equivalent cell sequences or paths along the structure starting and ending in a joining locus.

\begin{defi}
A {\bf generatex} $\mathcal{G}=(\mathcal{K},\mathcal{C})$ is a subtemplex in $T=(K,G)$, where $\mathcal{C}$ is a cycle of the digraph $G$, and $\mathcal{K} \subset K$. A generatex is said to be of order $\theta$ with $\theta \in \mathbb{N}$, $\theta \ge 1$, if the cycle $\mathcal{C}$ has $\theta$ ingoing nodes. In this case, the cycle $\mathcal{C}$ is said to be of order $\theta$ too. 
\end{defi}

\noindent
As strips in a template are always simple --- in the sense that they are defined so that they do not pass through the set of joining loci more than once --- we can introduce the definition of stripex as the generalization of a strip in a template. 

\begin{defi}
Let $\mathcal{G}_i$ be an order-$\theta$ generatex of $T=(K,G)$. 
If $\theta=1$, the {\bf stripex} $\mathcal{S}_i=\mathcal{G}_i$. 
If $\theta > 1$, $\mathcal{G}_i$ can be splitted into $\theta$ stripexes $\mathcal{S}_{i_j}$ with $1 \le j \le \theta$.  
\end{defi}

\noindent
Last but not least, a stripex will be said to have a local twist if  the free boundaries change their relative positions with respect to the orientation from the center to the periphery. Twisted stripexes correspond, in the dissipative 3D limit, to strips with uneven local torsions. From the point of view of an analysis in a Poincaré section, only the parity of the local torsion matters, and this intrinsic property can be directly extracted from the cell complex. If a finer characterization is required, it may be necessary to count the number of half-turns, as done with templates for three-dimensional cases.

\subsection{Dynamical units and taxonomy of chaos}
\label{dynit}

In his first paper on chaos,\cite{Ros76a} Otto Rössler described chaos as the combination of a two-variable oscillator with a one-variable nonlinear switch. These two ingredients can be considered as the fundamental units to build any type of chaos. As a fundamental unit, a periodic oscillator is characterized by a largest  Lyapunov exponent which is null. It can be furthermore either dissipative or conservative. An S-unit is a switch between two manifolds allowing stretching and squeezing, the process required for mixing trajectories and inducing the sensitivity to initial conditions.\cite{Gil98} It is by combining an O-unit with an S-unit that Rössler got his first chaotic system,\cite{Ros76a} which was only slightly more complicated than the popular Rössler system.\cite{Ros76c} 

After having proposed various systems for various types of chaos,\cite{Ros76b,Ros77a,Ros77c,Ros77d,Ros78} Rössler proposed to construct a chaos hierarchy \cite{Ros78b,Ros83} based on Lyapunov exponents. Such a hierarchy was then updated,\cite{Kle91,Let20b} before being transformed into a taxonomy of chaos,\cite{Let22a} for defining and classifying classes of chaotic invariant sets based on shared characteristics. The taxonomic ranks, from the rough to the most refined, are 
as follows.\cite{Let22a}
\begin{itemize}
\item dynamical or not;
\item stochastic or deterministic;
\item positive Lyapunov exponents;
\item negative Lyapunov exponents;
\item bounding tori;
\item modality of the first-return map;
\item template or templex.
\end{itemize}
As mentioned before, the numbers $p$ of positive and $q$ of negative Lyapunov 
exponents may be related to the existence of S- and O-units, respectively. This is used as guidelines for the present work, but further studies are required to predict the sign of the Lyapunov exponents from the O- and S-units.

Bounding tori are defined as permeable surfaces which bound chaotic 
attractors.\cite{Tsa04} For instance, the spiral Rössler attractor is bounded 
by a genus-1 torus, and the Lorenz attractor by a genus-3 torus. The genus $g$ 
of bounding tori indicates the minimum number $N_{\rm c}$ of components the 
Poincaré section must have: $N_{\rm c} = g - 1$ for $g \geqslant 3$ and 
$N_{\rm c} = 1$ for $g = 1$.\cite{Tsa04,Let22a} Here, from a templex in a 
reduced form as we shall explain, we can work at the level of the first-return map (or Poincaré section), that is, at the penultimate taxonomic rank which corresponds already to a very refined level of description.

O-units will be associated with a two-dimensional oscillator (the Lotka 
system,\cite{Lot10} the van der Pol equation,\cite{vdP26} the 
Brusselator,\cite{Pri68} etc.) producing a period-1 orbit. S-units --- without 
which there is no sensitivity to initial conditions, that is, chaos --- cannot 
exist without an O-unit: they must be bonded to an O-unit, as later described. The cycles computed from the reduced digraph will lead us to the decomposition of reduced templex into such fundamental units, as we shall see in the next section. 

\section{Towards dynamical units}
\label{dynamit}

The recipe to extract the fundamental units from a templex comprises two parts: reducing the templex, and decomposing its reduced directed graph into O- and S-units. A graph decomposition is defined as follows.\cite{Kot24}

\begin{defi}
\label{def:dec}
A {\bf{decomposition}} of a graph $G=(N,E)$ is such that:

\begin{itemize}
    \item $G=\bigcup_{1 \le i \le k}G_{i}$, where $G_i \subset G$, $\forall i$.
    \item $E(G_i) \cap E(G_j)= \emptyset$ $\forall i,j$.
\end{itemize} 
\end{defi}
We will here present how to proceed when the system is chaotic. The case of regular dynamics is discussed in Appendix \ref{regdyn}. The starting point is a generating templex. 

\begin{defi}
A {\bf generating templex} $T = (K,G)$ is a templex composed by a flow-oriented {\sc BraMAH} complex associated with a digraph such that there is only one outgoing 2-cell per component of the joining locus.  
\end{defi}

\noindent
Complex $K$ is thus assumed to have a single cell outgoing from each component 
of the joining locus.\cite{Cha22} The splitting locus in a templex $T$ is the analog of the splitting chart in a template. It is formed by the cells that do not belong to the joining locus and from which the flow can visit at least two non-equivalent stripexes. We will also assume that a given $\kappa_{\rm l}$-cell cannot be bounded by a joining locus and a splitting locus, so that there are at least two $\kappa_{\rm l}$-cells between a joining locus and a splitting locus. All the non-trivial dynamical processes will thus be captured between splitting and joining loci. The digraphs for our examples of joining and splitting loci are drawn in Figs\ \ref{splitjoin}(c) and \ref{splitjoin}(d). We define joining and splitting nodes as follows. 

\begin{defi}
\label{joinode}
A {\bf joining node} ({\bf splitting node}) with an indegree (outdegree) $\geq 2$ in a digraph $G=(N,E)$ corresponds to a cell that is adjacent to a joining (splitting) locus in the  complex. 
\end{defi}
Joining and splitting nodes cannot be removed during a merging process. Notice that joining and splitting nodes must be necessarily distinct, that is, a node cannot be simultaneously joining and splitting. By convention, a joining (splitting) node will be underlined (overlined). An implementation in Python of the reduction procedure detailed below can be found in {\url{https://git.cima.fcen.uba.ar/caterina.mosto/templex_properties_python}}.

The rationale of the reduction procedure consists in merging as many cells as possible in complex $K$ (and its corresponding nodes in digraph $G$) without altering the topological properties (homology and torsion groups, joining and splitting loci, set of generatexes or stripexes).  In other words, we can merge cells (nodes) as long as the operation does not create or eliminate stripexes, torsions, or holes. Once the complex is reduced ($\overline{K}$), it remains to decompose the reduced digraph $\overline{G}$ into O- and S-units. This procedure requires identifying the set of bonds ${\cal B}$ in a templex. 

\begin{defi}
Let  $\overline{T}=(\overline{K},\overline{G})$ be a reduced templex with $\overline{G}=(\overline{N},\overline{E})$. A {\bf bond} $B_i \in \mathcal{B}$ is such that $B_i \in \overline{E}$ and connects a joining node to a splitting node.
 \end{defi}
Bonds correspond to a part of the templex where the dynamics is trivial. In other words, the  specificity of the attractor occurs between the splitting and the joining  loci, that is, in the complementary part of the bond. Bonds can be thought of as the handle(s) of a template, holding together the stripexes and marking the paths where the dynamics is irrelevant. When there are $m$ bonds ($m \in \N_{\ge 2}$) belonging to a single cycle in $\overline{G}$, the bond is said to be degenerated and of order-$m$.

Once the bonds are identified and assembled in $\mathcal{B}$, we are ready to decompose the graph into units. We will start by selecting the O-units among the cycles of the reduced digraph. By construction, a cycle $i$ in $\overline{G}$ surrounds a center of rotation $c_i$, and is related to a bond (or set of degenerated bonds) $B_i$. 

\begin{defi}
The {\bf O-unit} associated with a bond (or with a set of degenerate bonds) is the cycle in $\overline{G}$ (i) having the minimum number of nodes and (ii) lying closest to a center of rotation $c_i$. 
 \end{defi}
There is just one O-unit per bond (or per set of degenerated bonds). If a bond is degenerated, the associated O-unit will also be considered as degenerated.

Once the O-units are identified, the S-units must be selected from the remaining cycles in $\overline{G}$ so that the set of units forms a decomposition of $\overline{G}$.

\begin{defi}
An {\bf S-unit} is a path obtained by subtracting the edges of the O-units from the remaining cycles in $\overline{G}$. 
 \end{defi}
Removing the edges corresponding to the O-units from the remaining cycles associated with the S-units exhibits that S-units cannot exist without the related O-unit. Notice that, in the absence of degenerated bonds, the total number of units (of both types) in the decomposition of $\overline{G}$ will typically coincide with the number of elements in the stripex set.

\section{Dynamical units for some chaotic attractors}
\label{example}

In this section, we provide the reduced templexes and their decomposition into O- and S-units for all the attractors which were investigated in Ref. \onlinecite{Cha22}. The Deng toroidal chaos is also treated.

\subsection{The Lorenz attractor}

Let us now consider the Lorenz system\cite{Lor63}
\begin{equation}
  \label{loreq63}
  \left\{
    \begin{array}{l}
      \dot{x} = - \sigma x + \sigma y \\[0.1cm]
      \dot{y} = Rx -y -xz \\[0.1cm]
      \dot{z} = -b z +xy
    \end{array}
  \right.
\end{equation}
with parameter values as selected by Lorenz, that is, $R = 28$, $\sigma = 10$, and $b = \frac{8}{3}$. The system is equivariant under a rotation symmetry ${\cal R}_z (\pi)$ \cite{Let01} and produces the Lorenz attractor which is bounded by a genus-3 torus.\cite{Tsa04} 

\begin{figure}[htbp]
  \centering
   \begin{tabular}{c} 
     \includegraphics[width=0.35\textwidth]{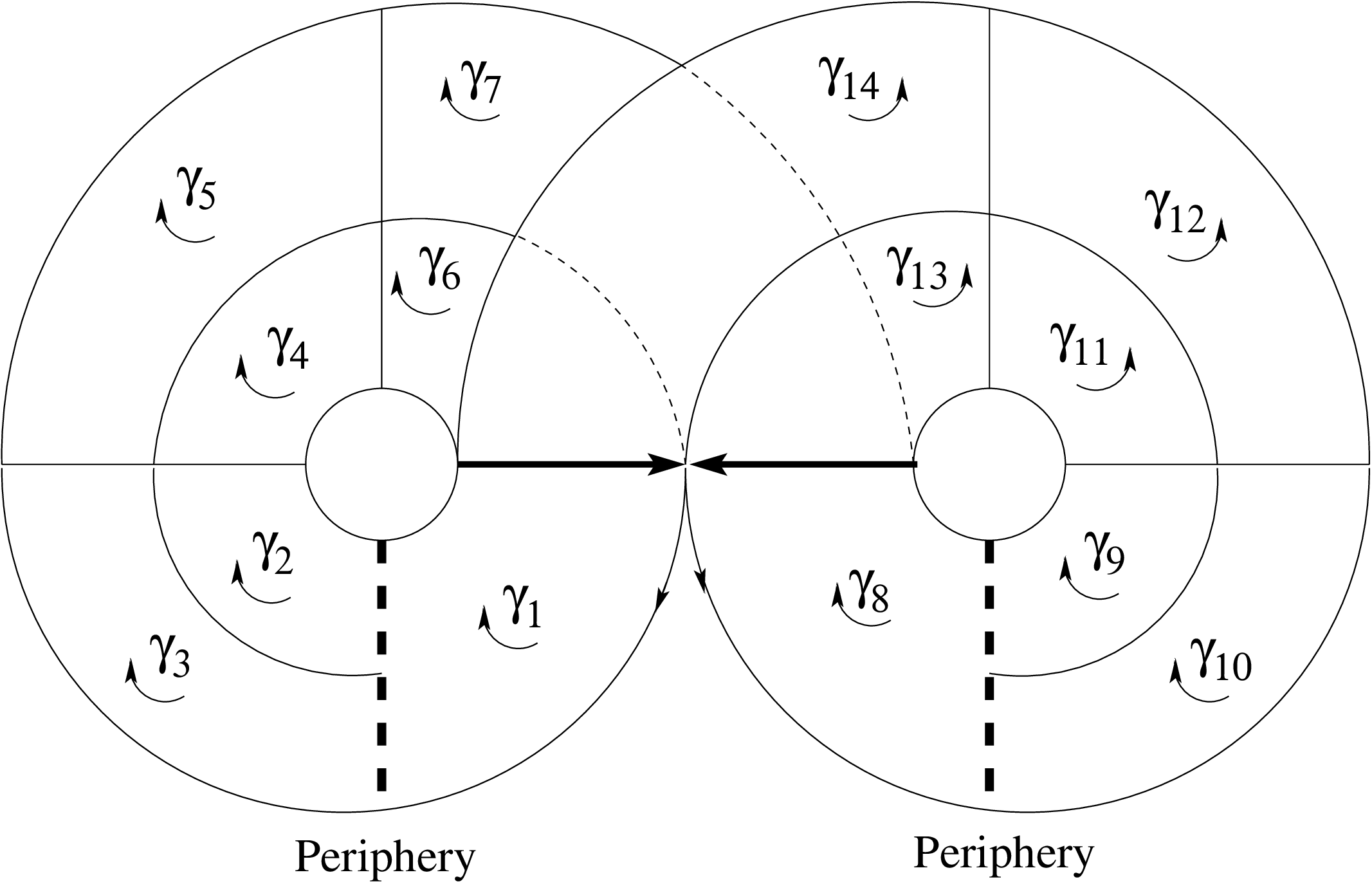} \\
     {\small (a) Semi-planar representation of generating complex $K_{\rm L}$} \\[0.2cm]
     \includegraphics[width=0.35\textwidth]{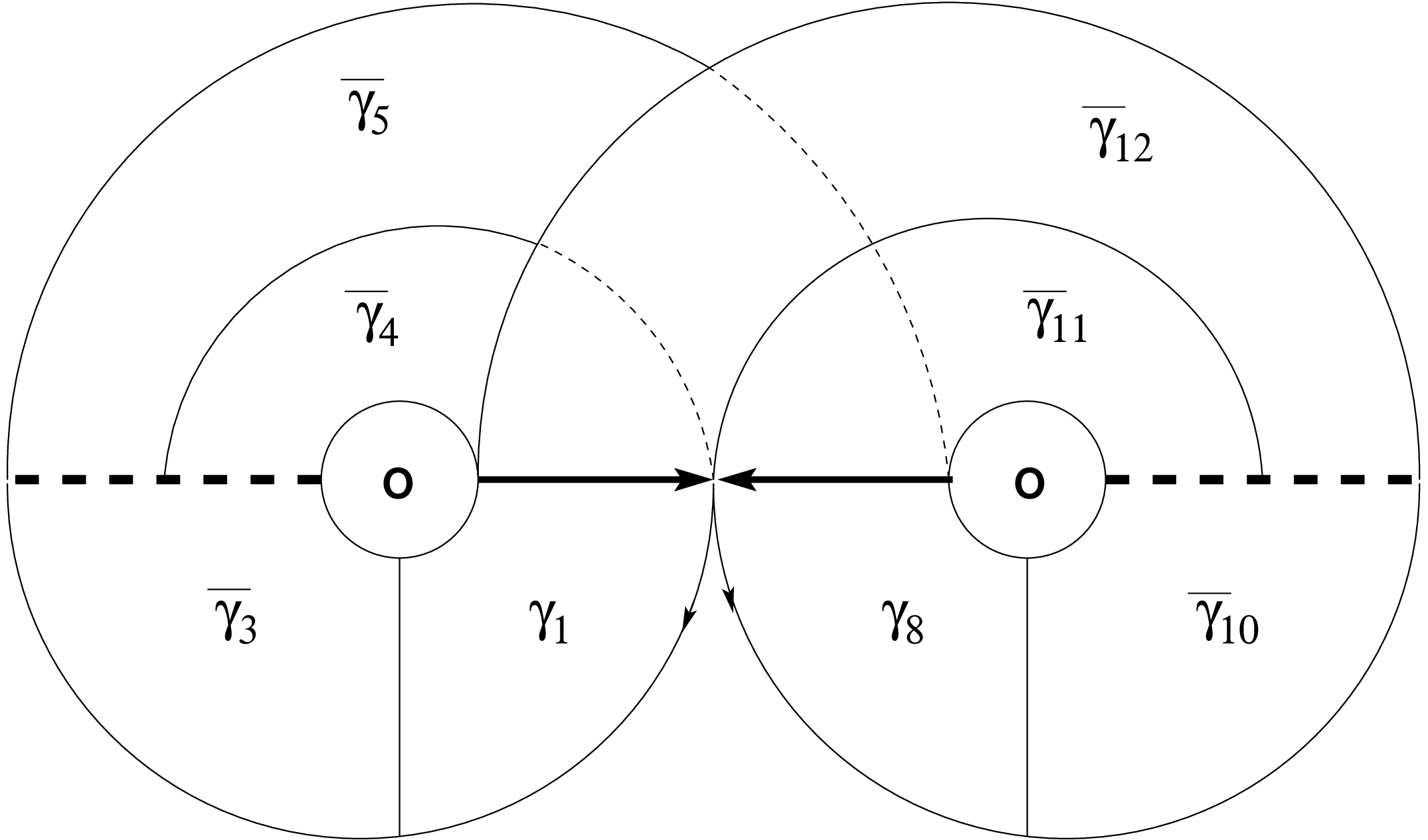} \\
     {\small (b) Semi-planar representation of reduced complex $\overline{K}_{\rm L}$} \\[0.2cm]
     \includegraphics[width=0.35\textwidth]{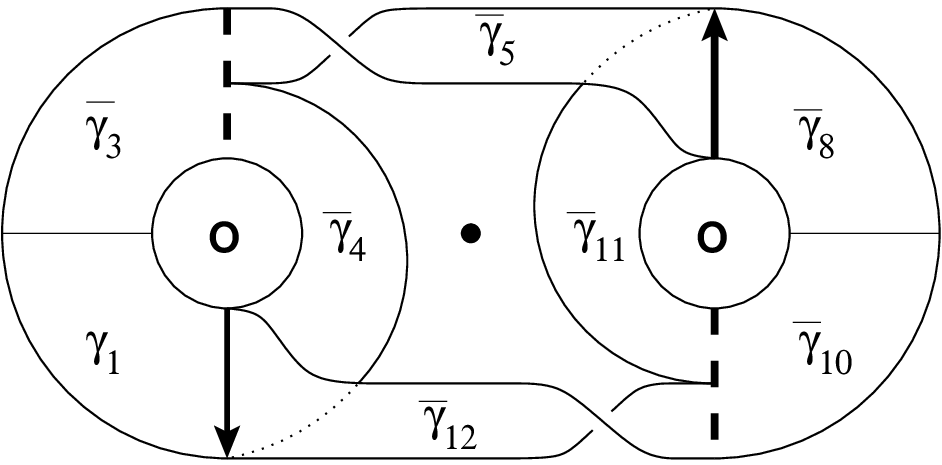}  \\[0.1cm]
     {\small (c) Lorenz template with cells of $\overline{K}_{\rm L}$} \\[0.2cm]
     \includegraphics[width=0.16\textwidth]{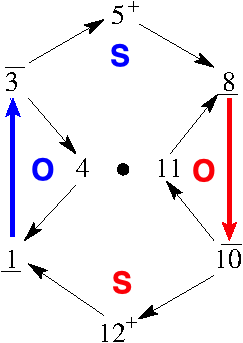} \\
     {\small (d) Reduced digraph $\overline{G}_{\rm L}$} with bonds in 
	   color.\\[-0.2cm]
  \end{tabular}
  \caption{Original generating complex $K_{\rm L}$ (a) from Ref. 
\onlinecite{Cha22} with oriented components of the joining locus as thick arrows and components of the splitting locus as dashed lines; reduced cell  complex $\overline{K}_{\rm L}$ (b); the template of the Lorenz attractor in which cells of $\overline{K}_{\rm L}$  are reported (c); reduced digraph $\overline{G}_{\rm L}$ with bonds in color (d).
	}
  \label{genelor}
\end{figure}

Fig.\ \ref{genelor}(a) shows the generating complex $K_{\rm L}$ leading to the 
templex $T_{\rm L} = (K_{\rm L},G_{\rm L})$ for the Lorenz attractor previously provided in Ref. \onlinecite{Cha22}. Complex $K_{\rm L}$ has a single $0$-generator since the attractor has no disconnected components: $H_0[K_{\rm L}] \sim \Z$. Since all the attractors here have a single connected component, we will not provide the $0$th-order homology group in the subsequent part of this paper. Complex $K_{\rm L}$ has two $1$-generators, one around the center of rotation in each wing of the Lorenz attractor ($H_1 [K_{\rm L}] \sim \Z^2$) and no enclosed cavities ($H_2 [K_{\rm L}] \sim 0$).  The Lorenz attractor has a joining locus made of two components, thus corresponding to a two-components Poincaré section.

Let us undertake the templex reduction. In $K_{\rm L}$ [Fig.~\ref{genelor}(a)], cells $\gamma_1$ and $\gamma_8$ are each bounded simultaneously by the joining locus and splitting locus. To prevent this, cells $\gamma_2$ and $\gamma_3$ ($\gamma_9$ and $\gamma_{10}$) are merged into one cell $\overline{\gamma}_3$ ($\overline{\gamma}_{10}$). The templex can be further reduced without altering its properties: cells $\gamma_5$ and $\gamma_7$ ($\gamma_4$ and $\gamma_6$) can be merged into one cell $\overline{\gamma}_5$ ($\overline{\gamma}_4$); the same merging can be applied in the right wing, leading to the reduced complex $\overline{K}_{\rm L}$ [Fig.\ \ref{genelor}(b)]. The reduced templex $\overline{T}_{\rm L}$ has eight $2$-cells instead of fourteen, but retains the same homologies, and the same generatex set made of three elements, one being of order $2$. The stripex set has four elements: two of them have a local twist. Merging more cells would alter the templex properties. The reduction is thus accomplished. 

The reduced complex $\overline{K}_{\rm L}$ can be drawn in a semi-planar representation [Fig.\ \ref{genelor}(b)], as well as in a template-like representation. It is possible to report in a template the cells of the reduced cell complex as drawn in Fig.\ref{genelor}(c), so that the correspondence between template and templex is clear. The template representation serves the purpose of displaying the genus-$3$ surface of the bounding torus. Notice that the two focus holes in the bounding torus correspond to the two $1$-holes exhibited in the cell complex, while the saddle hole around the rotation axis of the Lorenz symmetry in the bounding torus does not have the property required by a $1$-hole as considered with homologies.

The reduced digraph $\overline{G}_{\rm L}$ is shown in [Fig.\ \ref{genelor}(d)] with the point representing the rotation axis.\cite{Let01} A reduced digraph is typically formed by joining nodes, splitting nodes and a set of `regular' nodes, the latter ones being found in the path from a splitting node to a joining node. The eight-node reduced digraph for the Lorenz attractor has a joining locus made of two components (exhibited by the two corresponding cells which are underlined, a two-component splitting locus (the corresponding cells are overlined). The digraph has four regular cells. In the reduced digraph, we dressed regular nodes $5$ and $12$ with the superscript ``$^+$'' to indicate the presence of a positive half-twist, corresponding to the local twist in the associated stripexes. 

There are two bonds: $\mathcal{B}_{\rm L}=\{ B_1, B_2 \}$, where $B_1 = (\underline{1},\overline{3})$ and $B_2 = ( \underline{8},\overline{10})$ in $\overline{G}_{\rm L}$. They are highlighted in colors in Fig.\ \ref{genelor}(d). 

We are now in a position to decompose $\overline{G}_{\rm L}$ into units. We have three cycles:
\[
  \begin{array}{lcr}
     \mathcal{C}_1(\overline{G}_{\rm L}) 
	  = \underline{1} \rightarrow \overline{3} \rightarrow 4 
	  \rightarrow \underline{1} \\[0.1cm] 
   \mathcal{C}_2(\overline{G}_{\rm L})
	  = \underline{8} \rightarrow \overline{10} \rightarrow 11  
	  \rightarrow \underline{8} \\[0.1cm]
   \mathcal{C}_3(\overline{G}_{\rm L})
	  = \underline{1} \rightarrow \overline{3} \rightarrow 5^+ 
	  \rightarrow \underline{8} \rightarrow \overline{10} 
	  \rightarrow 12^+ \rightarrow \underline{1} \, . 
  \end{array}
\]
The first two cycles are of order-$1$, and the third one is of order-$2$. Cycles $\mathcal{C}_1(\overline{G}_{\rm L})$ and $\mathcal{C}_2(\overline{G}_{\rm L})$ share bond $B_1$; and cycles $\mathcal{C}_2(\overline{G}_{\rm L})$ and $\mathcal{C}_3(\overline{G}_{\rm L})$ share bond $B_2$, one being the symmetric of the other under the rotation symmetry. 

$\mathcal{C}_1(\overline{G}_{\rm L})$ associated with $B_1$ and $\mathcal{C}_2(\overline{G}_{\rm L})$ associated with $B_2$ are cycles having the minimum number of nodes (three), and lying closer to the rotation centers $c_1$ and $c_2$ of each wing of the attractor. They both qualify to be O-units.

Now that the O-units have been determined, let us remove the edges already present in the  O-units from the remaining cycle $\mathcal{C}_3(\overline{G}_{\rm L})$. This leads to the following disjoint paths:

\[
  \begin{array}{lcr}
    P_1 (\overline{G}_{\rm L}) 
	  = \overline{3} \rightarrow 5 \rightarrow \underline{8} \\[0.1cm] 
    P_2 (\overline{G}_{\rm L}) 
	  = \overline{10} \rightarrow 12 \rightarrow \underline{1} \, ,
  \end{array}
\]
defining the two paths that are qualified to S-units. Digraph $\overline{G}_{\rm L}$ has thus been decomposed into four dynamical units: two O-units and two S-units.

Note that the reduced digraph is decomposed in a way which is left invariant under the rotation symmetry. Note also that the O-units correspond to the stripexes respectively associated with the left and right wings of the attractor:
\[
  \begin{array}{lcr}
	  {\cal S}_1 (\overline{T}_{\rm L})
	  = \gamma_1 - \gamma_{3} - \gamma_4, 
	  \\[0.1cm] 
	  {\cal S}_2 (\overline{T}_{\rm L}) 
	  = \gamma_{8} - \gamma_{10} - \gamma_{11},
  \end{array}
\]
while the S-units are parts of the two stripexes forming the generatex of order $2$:
\[
  \begin{array}{lcr}
	  {\cal S}_3 (\overline{T}_{\rm L})
	  = \gamma_{1} - \gamma_{3} - \gamma_{5} 
	  - \gamma_{8} \\[0.1cm] 
	  {\cal S}_4 (\overline{T}_{\rm L}) 
	  = \gamma_{8} - \gamma_{10} - \gamma_{12}
	 - \gamma_{1} \, ,
  \end{array}
\]

The result is consistent with the Lorenz attractor being a C$^1$T$^1$ chaos with a structure O-S$\vdots$S-O where $\vdots$ designates the transition from one wing to the other; this dynamical feature is produced by the S-units.

\subsection{The spiral and funnel Rössler attractors}
\label{rosatts}

The Rössler attractors are produced by one of the most common chaotic systems proposed by Otto E. Rössler, reading as\cite{Ros76c}
\begin{equation}
  \left\{
    \begin{array}{l}
      \dot{x} = -y - z \\[0.1cm]
      \dot{y} = x + ay \\[0.1cm]
      \dot{z} = b + - cz + xz \, . 
    \end{array}
  \right.
\end{equation}
Setting $b = 2$, and $c = 4$, a spiral attractor is obtained for $a = 0.432$ and a funnel attractor for $a = 0.492$.\cite{Let95a} A spiral Rössler attractor is characterized by $p = 1$ positive and $q = 1$ negative Lyapunov exponents: it is a C$^1$T$^1$ attractor. 

\begin{figure}[ht]
  \begin{tabular}{cc}
    \includegraphics[width=0.24\textwidth]{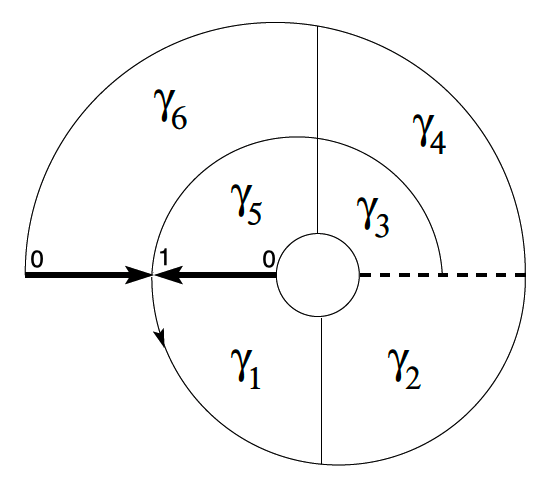} &
    \includegraphics[width=0.24\textwidth]{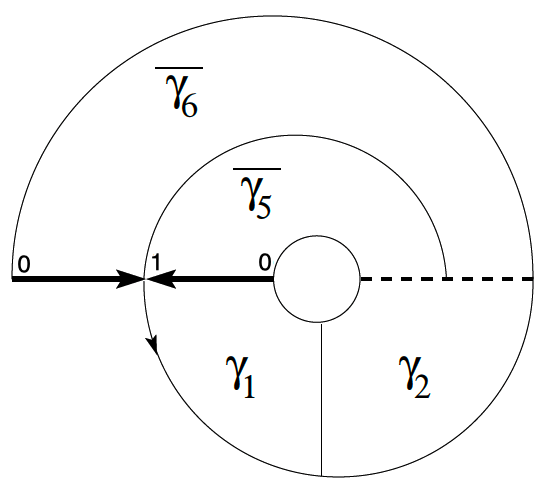} \\
    {\small (a) Complex $K_{\rm R_s}$} &
    {\small (b) Reduced complex $\overline{K}_{\rm R_s}$} \\[0.2cm]
    \multicolumn{2}{c}{
      \includegraphics[width=0.30\textwidth]{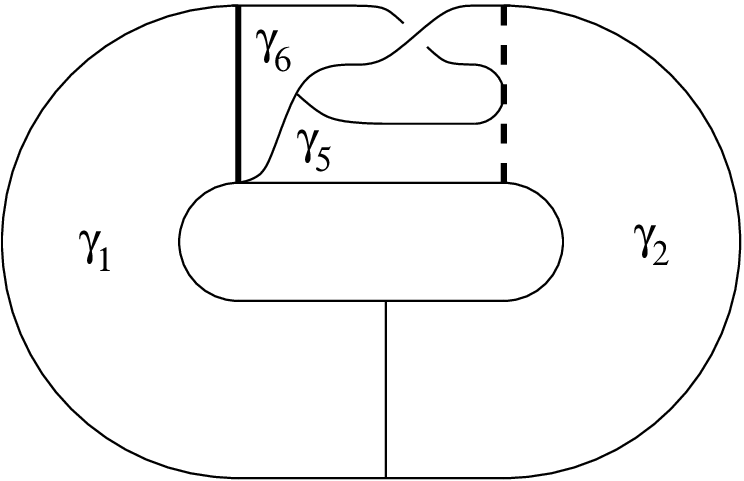}} \\
    \multicolumn{2}{c}{\small (c) Template with the cells of $\overline{K}_{\rm R_s}$} \\[0.2cm]
    \multicolumn{2}{c}{
      \includegraphics[width=0.14\textwidth]{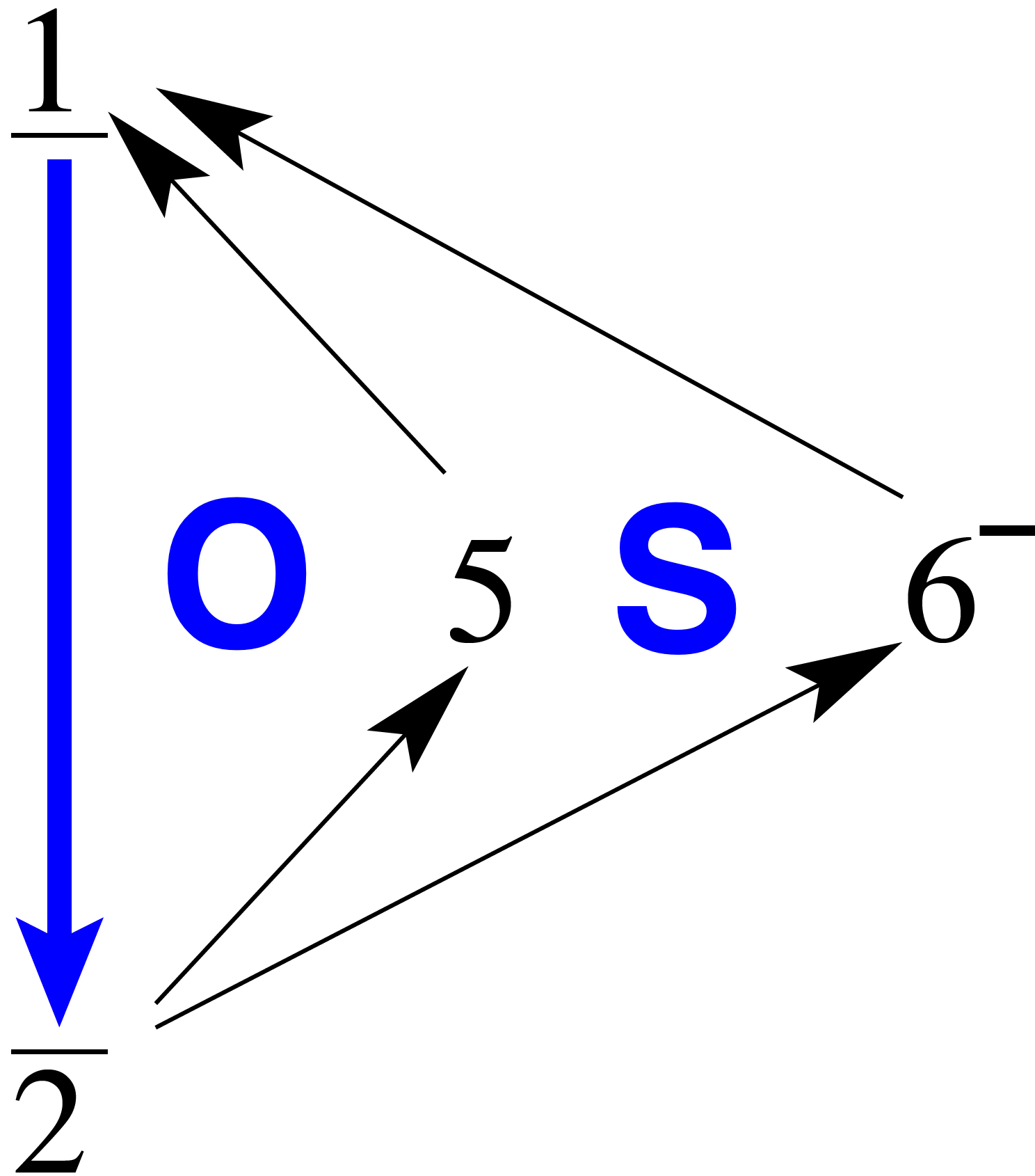}} \\
    \multicolumn{2}{c}{\small (d) Reduced digraph $\overline{G}_{\rm R_s}$ with 
	  bond as a blue arrow} 
	  \\[-0.2cm]
  \end{tabular}
  \caption{Generating complex $K_{\rm R_s}$ from Fig. 12 in Ref. \onlinecite{Cha22} for the spiral Rössler attractor and its reduced version $\overline{K}_{\rm R_s}$ (b). The splitting lines are dashed and the joining locus are thick. The corresponding reduced  digraph $\overline{G}_{\rm R_s}$ with its dynamical units is drawn in (d). The bond $\underline{1} \rightarrow \overline{2}$ is colored 
	in blue.  }
  \label{spidigra}
\end{figure}

Fig.\ \ref{spidigra}(a) shows the generating complex $K_{\rm R_s}$ for the spiral Rössler templex taken from Fig.\ 12 in Ref. \onlinecite{Cha22}. The corresponding reduced complex  $\overline{K}_{\rm R_s}$ is drawn in Fig.\ \ref{spidigra}(b). In terms of homologies, the complex $K_{\rm R_s}$ has one $1$-generator ($H_1 [K_{\rm R_s}] \sim \Z$) surrounding the saddle-focus singular point: there are no enclosed cavities ($H_2 [K_{\rm R_s}] \sim 0$). It has a one-component joining locus: the thick black arrows in Fig.\ \ref{spidigra}(a) and (b). The joining locus appears duplicated because $K_{\rm R_s}$ is drawn as an unfolded planar object. In planar representations of cell complexes, repeated $0$-cells indicate gluing prescriptions. There are two stripexes for the spiral Rössler templex, one of which has a local twist, corresponding to the negative half-twist in the 
template of Fig.\ \ref{spidigra}(c).\cite{Let95a} This is indicated with the 
superscript ``$^-$'' in node $6$. 

The reduced digraph for the spiral R\"ossler attractor $\overline{G}_{\rm R_s}$ is shown in Fig.\ \ref{spidigra}(d). It has two order-1 cycles 
\[
  \begin{array}{l}
	\mathcal{C}_1(\overline{G}_{\rm R_s})= \underline{1} \rightarrow \overline{2} \rightarrow 5 \rightarrow 1 \\[0.1cm] 
	\mathcal{C}_2(\overline{G}_{\rm R_s})= \underline{1} \rightarrow \overline{2} \rightarrow 6 \rightarrow \underline{1}
  \end{array}
\]
There is a single bond $\mathcal{B}_{\rm R_s}= \{(\underline{1}, \overline{2}) \}$, drawn as a blue arrow in Fig.\ \ref{spidigra}(d). The two cycles have the same number of nodes, but the one lying closer to the center of rotation is $\mathcal{C}_1(\overline{G}_{\rm R_s})$, identified hence as the O-Unit. Removing the bond from the remaining cycle $\mathcal{C}_2(\overline{G}_{\rm R_s})$ leads to the S-unit, described by the path $\overline{2} \rightarrow 6 \rightarrow \underline{1}$. 

As before, the number of units corresponds to the number of  stripexes (or strips in the dissipative 3D limit), that is, to the number of monotone branches in the first-return map (not shown). The reduced digraph actually works at the same taxonomic rank as the first-return map.

\begin{figure}[ht]
  \begin{tabular}{c}
    \includegraphics[width=0.30\textwidth]{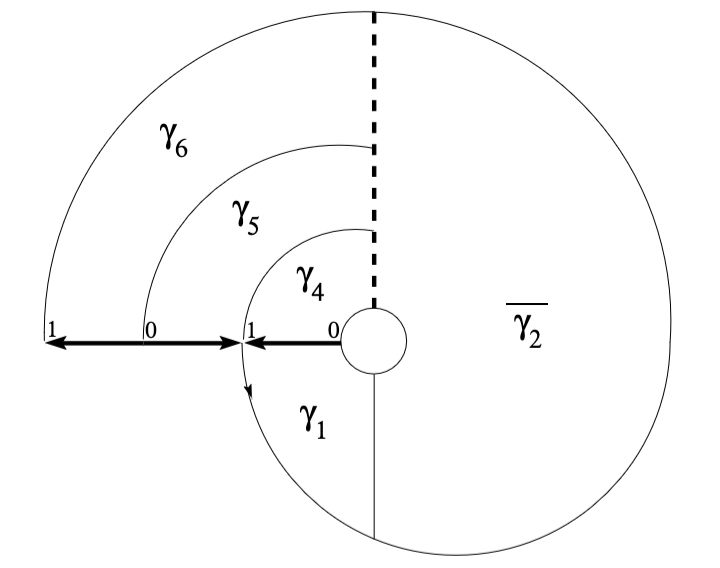} \\[0.1cm]
    {\small (a) Planar diagram of reduced complex $\overline{K}_{\rm R_f}$} 
	  \\[0.2cm]
    \includegraphics[width=0.38\textwidth]{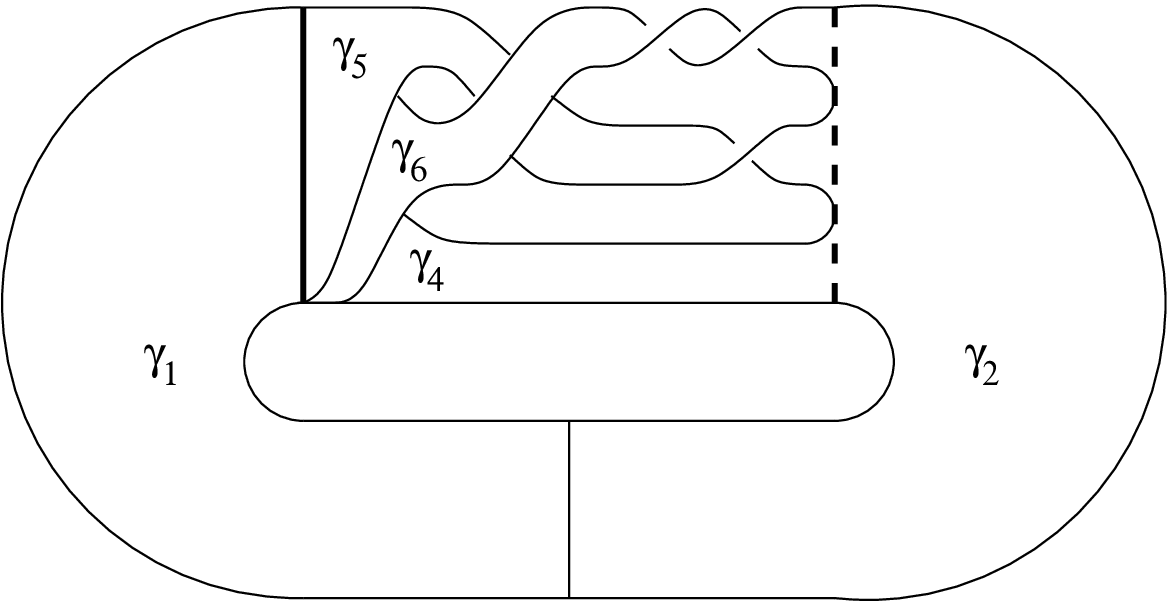} \\
	  {\small (b) Template \added{with the cells of $\overline{K}_{\rm R_f}$ }} \\[0.2cm]
    \includegraphics[width=0.22\textwidth]{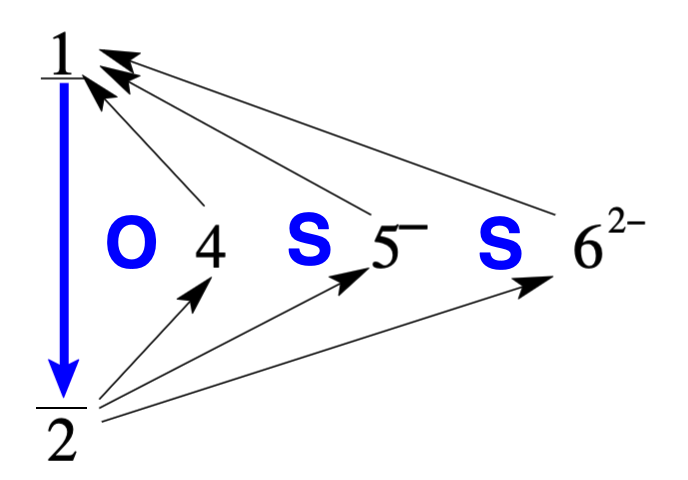} \\
    {\small (c) Reduced digraph $\overline{G}_{\rm R_f}$} with the bond 
	  as a blue arrow \\[-0.2cm]
  \end{tabular}
  \caption{Reduced templex for the funnel Rössler attractor $\overline{T}_{\rm R_f} = (\overline{K}_{\rm R_f}, \overline{G}_{\rm R_f})$ (corresponding to Fig. 15 in Ref. \onlinecite{Cha22}).}
  \label{fundigra}
\end{figure}

When $a = 0.492$, the first-return map to a Poincaré section has three branches, meaning that there is a third stripex (strip) in the templex (template).\cite{Let95a,Cha22} From Fig. 15 in Ref. \onlinecite{Cha22}, a reduced complex is obtained as shown in Fig. \ref{fundigra}(a). The cells are reported in the template as initially extracted by Letellier and coworkers\cite{Let95a} as shown in Fig. \ref{fundigra}(b). As expected, the complex $K_{\rm R_f}$ has the same homology groups as $K_{\rm R_f}$: a single $1$-generator around the focus-type hole of the attractor ($H_1 [K_{\rm R_f}] \sim \Z$) and no enclosed cavities ($H_2 [K_{\rm R_f}] \sim 0$). There is a one-component joining locus in the $1$-cell $\langle 0,1 \rangle$, which appears thrice (and not twice as before) in the planar diagram, showing that the gluing  prescription is more complex than for the spiral attractor. In the reduced 
digraph [Fig.\ \ref{fundigra}(c)], the three order-1 cycles are
\[
  \begin{array}{l}
    \mathcal{C}_1(\overline{G}_{\rm R_f}) = \underline{1} \rightarrow 
	  \overline{2} \rightarrow  4 \rightarrow \underline{1} \\[0.1cm] 
    \mathcal{C}_2(\overline{G}_{\rm R_f}) =  \underline{1} \rightarrow 
	  \overline{2}  \rightarrow 5 \rightarrow \underline{1} \\[0.1cm]
    \mathcal{C}_3(\overline{G}_{\rm R_f}) = \underline{1} \rightarrow 
	  \overline{2} \rightarrow 6 \rightarrow \underline{1}.
  \end{array}
\]
There is a single bond $\mathcal{B}_{\rm R_f}=\{ (\underline{1}, \overline{2}) \}$. The three cycles $\mathcal{C}_1, \mathcal{C}_2, \mathcal{C}_3$ have the same number of nodes, but $\mathcal{C}_1$ is closest to the rotation center. It is therefore identifed with the O-unit. The S-units are obtained by removing the bond (which is already in the O-unit) from the two remaining cycles, leading to $(\overline{2} \rightarrow 5 \rightarrow \underline{1})$ and $(\overline{2} \rightarrow 6 \rightarrow  \underline{1})$.

With respect to the spiral Rössler case, the funnel Rössler [Fig.\ \ref{fundigra}]  presents an additional S-unit. The two S-units are in ``parallel'' (as in an electric circuit), that is, they are bonded to the O-unit with the same path $(\underline{1} \rightarrow \overline{2})$. From Ref. \onlinecite{Cha22}, we know that stripex $\mathcal{S}_2$ (related to the cycle $\mathcal{C}_2$) has a local twist. The template strip corresponding to stripex $\mathcal{S}_3$ has a negative $2 \pi$-twist which is indicated with superscript ``$^{2-}$'' attributed to node $6$ in Fig.  \ref{fundigra}(c).

\subsection{The Burke and Shaw attractor}
\label{BSatt}

Another non-trivial example is provided by the Burke and Shaw attractor produced by the Burke and Shaw system\cite{Sha81}
\begin{equation}
  \label{Shaweq81}
  \left\{
    \begin{array}{l}
      \dot{x} = - a x - a y \\
      \dot{y} = - y - a xz \\
      \dot{z} = b + a xy 
    \end{array}
  \right.
\end{equation}
with parameter values as $a=10$, and $b = 4.271$. Integrating this system, a particular attractor is obtained, with a rotation symmetry ${\cal R}_z (\pi)$ and whose direct template is drawn in Ref. \onlinecite{Let96a}.  

\begin{figure}[ht]
  \centering
	\includegraphics[width=0.48\textwidth]{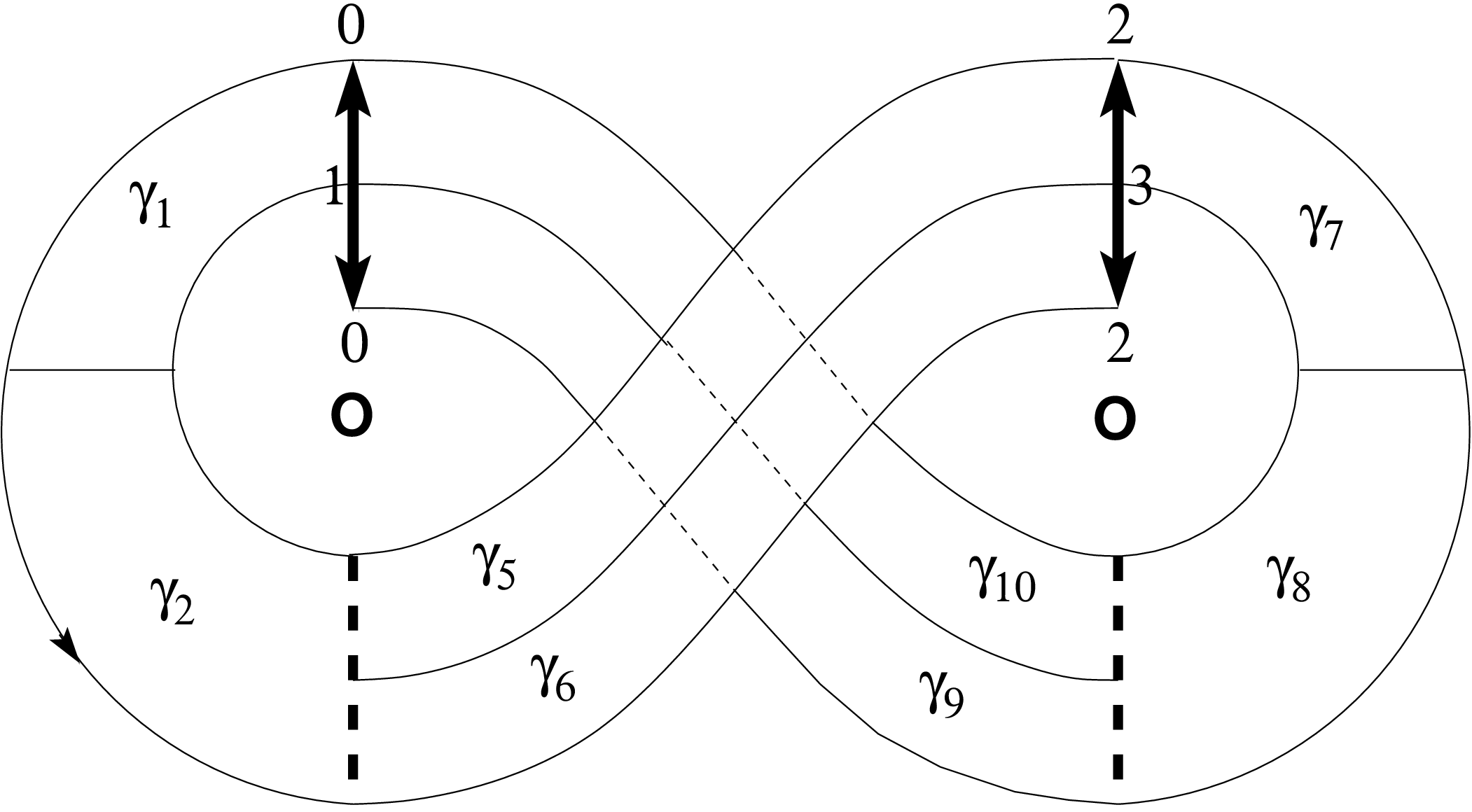}  \\[0.1cm]
	(a) Semi-planar reduced complex $\overline{K}_{\rm BS}$ \\[0.2cm]
  \includegraphics[width=0.33\textwidth]{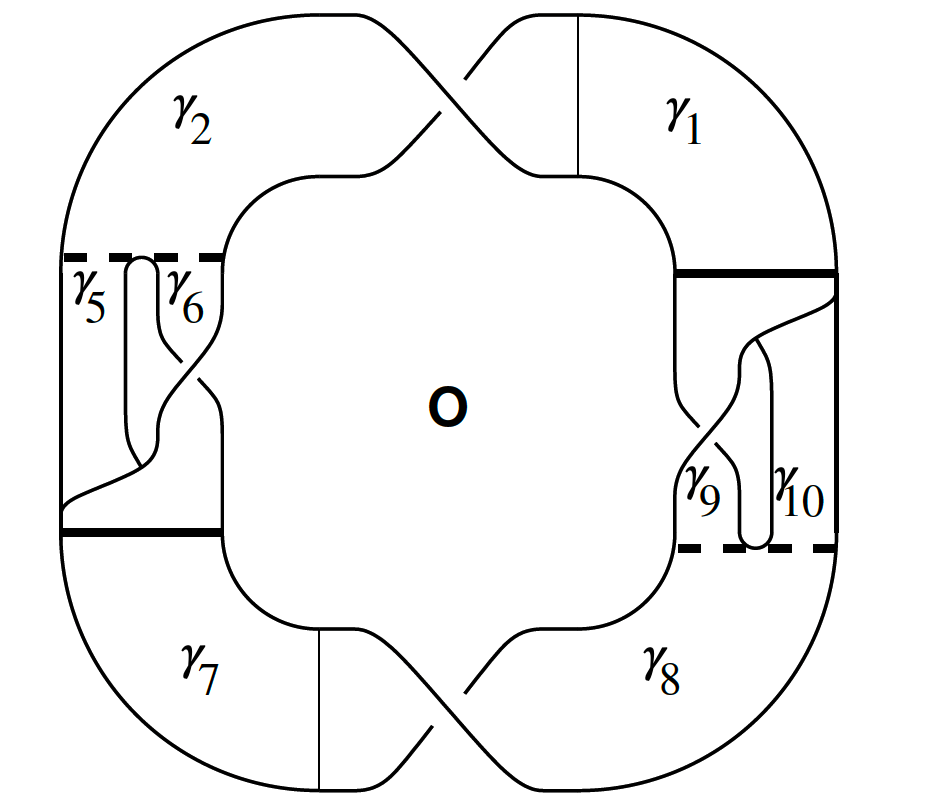}  \\
	(b) Burke Shaw template with the cells of $\overline{K}_{\rm BS}$ 
	\\[0.2cm]
  \includegraphics[width=0.30\textwidth]{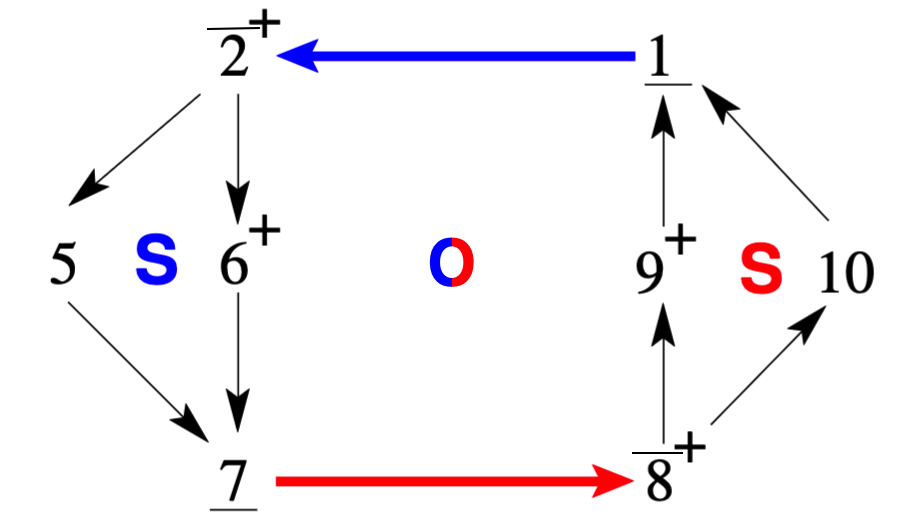} \\
	(c) Reduced digraph $\overline{G}_{\rm BS}$ with colored bonds and units\\[-0.1cm]
\caption{Reduced templex  $\overline{T}_{\rm BS} = (\overline{K}_{\rm BS}, \overline{G}_{\rm BS})$ for the Burke and Shaw attractor. $\overline{K}_{\rm BS}$ is shown in (a), followed by the template representation with the cell numbers on the strips (b). The reduced digraph $\overline{G}_{\rm BS}$ with the colored bond, and the fundamental units are shown in (c).}

  \label{fig:redBS}
\end{figure}

The cell complex $\overline{K}_{\rm BS}$ of the Burke and Shaw attractor has a single $1$-generator surrounding the focus-type hole in the attractor ($H_1 [K_{\rm BS}] \sim \Z$) and no enclosed cavities ($H_2 [K_{\rm BS}] \sim 0$). There is a joining locus with the two components $J_1({\rm BS})=\langle 1,0 \rangle$ and $J_2({\rm BS})=\langle 3,2 \rangle$, corresponding to the two components of the Poincaré section and associated to the two outgoing cells $\gamma_1$ and $\gamma_7$. As usual, joining loci in the complex are marked with thick arrows.

The reduced complex $\overline{K}_{\rm BS}$ for this attractor, obtained from the complex drawn in Fig. 19(b) from Ref. \onlinecite{Cha22}, is shown in a semi-planar representation in Fig.\ \ref{fig:redBS}(a). The direct template representation shown in Fig.\ \ref{fig:redBS}(b), has the two centers of rotation merged into a single one, illustrating how this attractor is actually bounded by a genus-one torus.\cite{Byr04} The reduced digraph $\overline{G}_{\rm BS}$ is shown in Fig.\ \ref{fig:redBS}(c). There are four cycles of order-$2$:
\[
  \begin{array}{l}
    \mathcal{C}_1(\overline{G}_{\rm BS})
	  = \underline{1} \rightarrow \overline{2} \rightarrow 6
	  \rightarrow \underline{7} \rightarrow 
	  \overline{8} \rightarrow 9 \rightarrow \underline{1}  \\[0.1cm]
    \mathcal{C}_2(\overline{G}_{\rm BS})
	  = \underline{1} \rightarrow \overline{2} \rightarrow 6
	  \rightarrow \underline{7} \rightarrow 
	  \overline{8} \rightarrow 10 \rightarrow \underline{1}  \\[0.1cm]
    \mathcal{C}_3(\overline{G}_{\rm BS})
	  = \underline{1} \rightarrow \overline{2} \rightarrow 5
	  \rightarrow \underline{7} \rightarrow 
	  \overline{8} \rightarrow 9 \rightarrow \underline{1}  \\[0.1cm]
    \mathcal{C}_4(\overline{G}_{\rm BS})
	  = \underline{1} \rightarrow \overline{2} \rightarrow 5
	  \rightarrow \underline{7} \rightarrow 
	  \overline{8} \rightarrow 10 \rightarrow \underline{1}  \, . 
  \end{array} 
\]
Strictly speaking, only the first three cycles are needed for all the nodes to be visited once. There are two bonds, namely, $\mathcal{B}_{\rm BS} = \{ B_1, B_2 \} = \{ (\underline{1}, \overline{2}), ~(\underline{7},\overline{8}) \}$. The four cycles have the same number of nodes, but the closest to the rotation center is the first one. Cycle $\mathcal{C}_1$ is thus an O-unit and it is degenerated since it crosses twice the joining locus. The degeneracy of this O-unit comes from the order-2 symmetry viewed as the union of two copies of an image as discussed in Refs. \onlinecite{Mir93,LetPhD,Let01} (a complete treatment of such an approach is currently under development). From the remaining cycles, we remove all the edges already present in the O-unit, and we are left with two S-units formed by  the paths 
\[
  \begin{array}{lcr}
    P_1 (\overline{G}_{\rm L}) 
	  = \overline{2} \rightarrow 5 \rightarrow \underline{7} \\[0.1cm] 
    P_2 (\overline{G}_{\rm L}) 
	  = \overline{8} \rightarrow 10 \rightarrow \underline{1} \, .
  \end{array}
\]
Let us recall that the four order-2 cycles lead to the following four stripexes 
\[
  \begin{array}{l}
    \mathcal{S}_1(\overline{T}_{\rm BS})= \gamma_{1} -
	  \gamma_{2} - \gamma_6 - \gamma_{7} \\[0.1cm]
    \mathcal{S}_2(\overline{T}_{\rm BS})=  \gamma_{7} - 
	  \gamma_{8} - \gamma_9 - \gamma_{1}  \\[0.1cm]
    \mathcal{S}_3(\overline{T}_{\rm BS}) = \gamma_{1} - 
	  \gamma_{2} - \gamma_5 - \gamma_{7} \\[0.1cm]
    \mathcal{S}_4(\overline{T}_{\rm BS}) = \gamma_{7} - 
	  \gamma_{8} - \gamma_{10}  - \gamma_{1}  
  \end{array} 
\]
where the union of the first two stripexes  forms the degenerated O-unit. If the Burke and Shaw attractor appears as the dual Lorenz attractor, its structure is quite different. Note that, from the theory of bounding tori, a single-component Poincaré section is suggested. The templex approach clearly exhibits the two-components of the joining locus, and so for the Poincaré section, in agreement with early developments for computing first-return maps to this attractor.\cite{Let96a,Byr04}

\subsection{A non-trivial four-dimensional attractor}

\begin{figure}[htbp]
  \centering
 \includegraphics[width=0.22\textwidth]{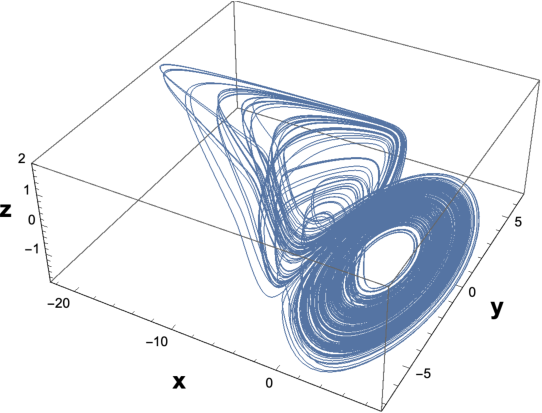}
    \includegraphics[width=0.22\textwidth]{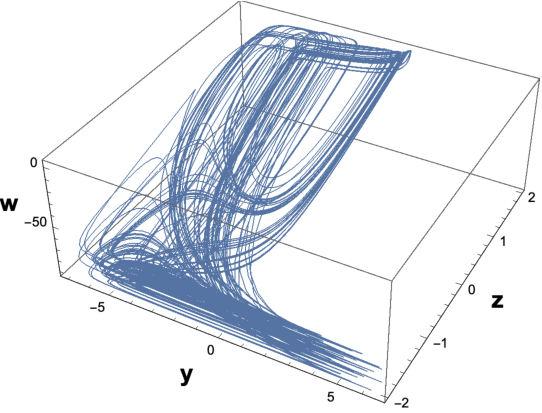}
 \caption{Two views of the chaotic attractor produced by the four-dimensional system (\ref{DMSeq}). Parameter values: $a = 7$, $b = 1.45$, $c = 1$, $d = 0.5$, $R = 6$, $\alpha = 0.3$, $\beta = 7$, $\epsilon_1 = 0.165$, $\epsilon_2 = 0.01$, $\epsilon_3 = 2$, and $\mu = 1.543$. }
  \label{litmonst}
  \centering
  \includegraphics[width=0.50\textwidth]{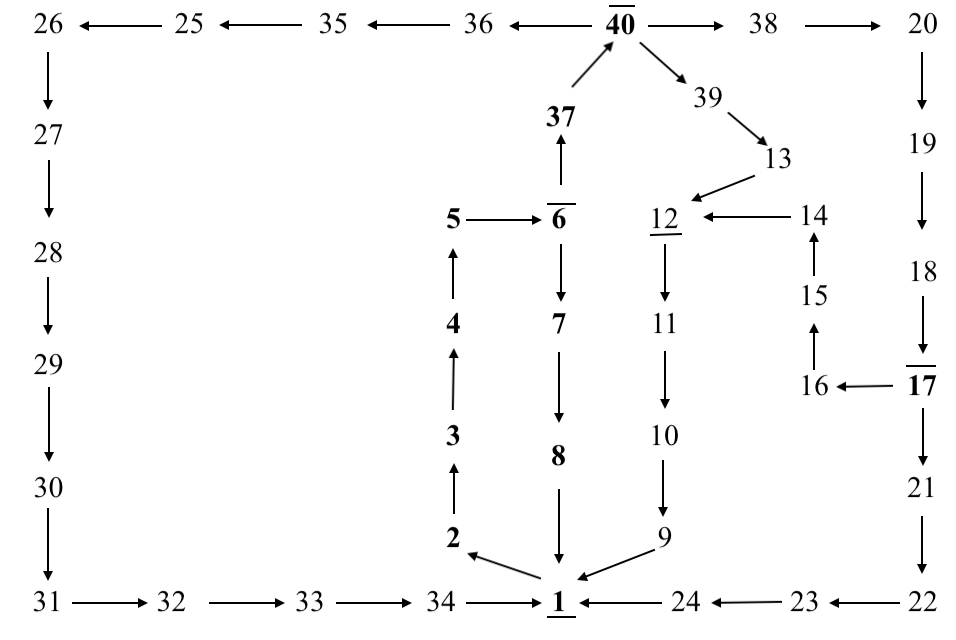} \\ 
  {\small (a) $G_{\rm S}$ in templex $T_{\rm S}$ from Ref. \onlinecite{Cha22}}. \\[0.2cm]
  \includegraphics[width=0.30\textwidth]{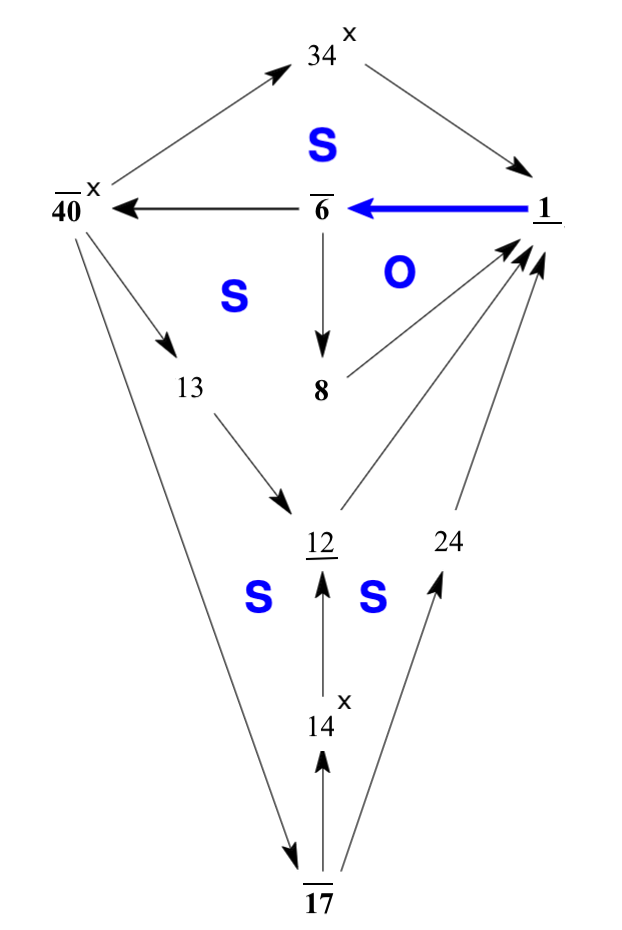} \\
	{\small (b) $\overline{G}_{\rm S}$ with colored bond, O- and S-units.} \\[-0.2cm]
  \caption{Digraph $G_{\rm S}$ (a) and reduced digraph $\overline{G}_{\rm S}$ (b) for the four-dimensional attractor produced by the system (\ref{DMSeq}). The bond is colored in blue.}
  \label{Digraph_4D}
\end{figure}

Ref.~\onlinecite{Cha22} computes the templex for a four-dimensional dynamical system, obtained as an extension of a three-dimensional one proposed by 
Deng.\cite{Den94} The system reads\cite{Sci01b} 
\begin{equation}
  \label{DMSeq}
  \left\{
    \begin{array}{l}
      \dot{x} = -(z+2) d
        \left( \displaystyle x - \left[ \displaystyle a + \epsilon_3 (2 + w)
            \right] \right) + (2-z) \\
            \hspace{0.5cm}\left[ \displaystyle \alpha (x-2) - \beta y - \alpha \, (x-2) \,
            \frac{\displaystyle (x-2)^2 +y^2}{R^2} \right]  \\[0.3cm]
            \displaystyle
      \dot{y} = -(z+2) \, (y-b) + (2-z) \\
         \hspace{0.5cm}\left[ \displaystyle \beta \, (x-2) + \alpha y
           - \alpha y \frac{\displaystyle (x-2)^2 + y^2 }{R^2} \right] \\[0.3cm]
            \displaystyle
      \dot{z} = (4-z^2) \, \frac{z + 2 - \mu (x+2)}{\epsilon_1} - c z \\[0.3cm]
            \displaystyle
      \dot{w} = (4-z^2) \, \frac{z + 2 - \mu (x+2)}{\epsilon_2} - c z \, .
    \end{array}
  \right.
\end{equation}
For the parameter values reported in the caption of Fig.\ \ref{litmonst}, this system produces a quite complicated attractor. Its cell complex $K_{\rm S}$ has five $1$-generators and there are no enclosed cavities. This reads $H_1 [K_{\rm S}] \sim \Z^5, H_2 [K_{\rm S}] \sim 0 $. The joining locus has two components.

The digraph $G_{\rm S} \in T_{\rm S}$ is drawn in Fig.\ \ref{Digraph_4D}(a). The reduced digraph $\overline{G}_{\rm S}$ is shown in Fig.\ \ref{Digraph_4D}(b).  We have five order-1 cycles
\[
  \begin{array}{lr}
    \mathcal{C}_1(\overline{G}_{\rm S}) 
	 = \underline{1} \rightarrow \overline{6} \rightarrow 8 \rightarrow 
	  \underline{1} \\[0.1cm] 
    \mathcal{C}_2(\overline{G}_{\rm S}) 
	  = \underline{1} \rightarrow \overline{6} \rightarrow \overline{40} 
	  \rightarrow 34 \rightarrow \underline{1}  \\[0.1cm]
    \mathcal{C}_3(\overline{G}_{\rm S}) 
	= \underline{1} \rightarrow \overline{6} \rightarrow \overline{40} 
	  \rightarrow 
	  13 \rightarrow \underline{12} \rightarrow \underline{1} \\[0.1cm]
    \mathcal{C}_4(\overline{G}_{\rm S}) 
        = \underline{1} \rightarrow \overline{6} \rightarrow \overline{40} 
	  \rightarrow 
	  \overline{17}  \rightarrow 24 \rightarrow \underline{1}  \\[0.1cm]
    \mathcal{C}_5(\overline{G}_{\rm S}) 
	  = \underline{1} \rightarrow \overline{6} \rightarrow \overline{40}
	  \rightarrow \overline{17} \rightarrow 14 \rightarrow \underline{12} 
	  \rightarrow \underline{1}  \, , 
  \end{array}
\]
and a single bond 
\[ \mathcal{B}(\overline{G}_{\rm S}) = \{ (\underline{1},\overline{6}) \} \, .
\]
Cycle $\mathcal{C}_1(\overline{G}_{\rm S})$ is the shortest (in number of nodes) and the closest to the rotation center: it corresponds therefore to the O-unit. Subtracting the edges of the O-unit from the remaining cycles we obtain the four S-units shown in Fig.~\ref{Digraph_4D}(b). There are as many units as stripexes in $T_{\rm S}$. Local twists are indicated with a $^\times$ superscript in nodes $14$, $34$ and $40$.

\begin{figure}[ht]
  \centering
	\includegraphics[width=0.30\textwidth]{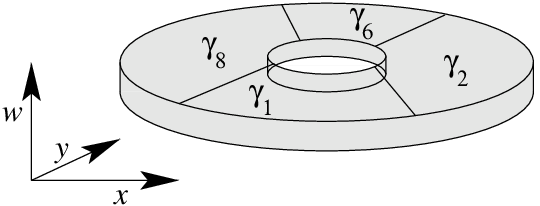} \\[-0.2cm]
  \caption{Filled torus corresponding to the O-unit of the four-dimensional 
chaotic attractor produced by the system (\ref{DMSeq}).} 
  \label{solidring}
\end{figure}

\begin{figure}[ht]
  \centering
  \includegraphics[width=0.30\textwidth]{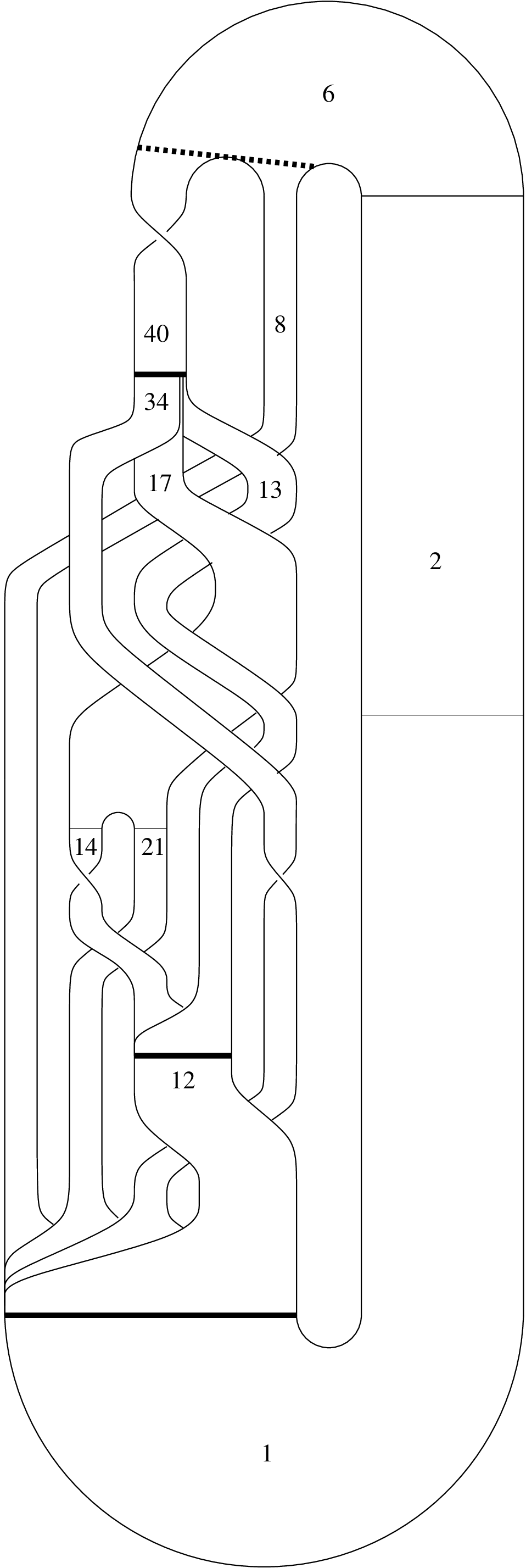} \\[-0.2cm]
  \caption{Direct template for the attractor produced by the four-dimensional system (\ref{DMSeq}), with the cells labeled as in reduced complex $\overline{K}_{\rm S}$. Splitting loci (dashed) and joining loci (thick) are drawn.}
  \label{dirtemons}
\end{figure}

Let us note that cells $\gamma_1$ to $\gamma_8$ and $\gamma_{40}$ are $3$-cells. The nodes corresponding to the three-dimensional cells are highlighted in bold in $G_{\rm S}$ and $\overline{G}_{\rm S}$. The O-unit is thus made of $3$-cells, approximating the solid torus or ring (see Fig.\ \ref{solidring}). The S-unit associated with $\mathcal{C}_2$ is particular since it is made of four nodes, due to the unusual splitting locus in the outgoing $3$-cell $\gamma_{40}$ (see Fig.\ \ref{dirtemons}). We need further investigations with other systems to determine whether this is a four-dimensional signature of the dynamics. The S-units associated with $\mathcal{C}_3$ and $\mathcal{C}_4$ contain $5$ nodes, and the S-unit corresponding to $\mathcal{C}_5$ contains $6$ nodes. The complexity of this attractor is clearly exhibited from these low-order cycles.

We were able to draw a template for this attractor as shown in Fig.~\ref{dirtemons}: according to the Birman and Williams theorem,\cite{Bir83a} this attractor must be C$^1$T$^1$ as confirmed by the four Lyapunov exponents (kindly computed for us by N. Stankevich): $\Lambda_1= 0.0469$, $\Lambda_2= 0.0$, $\Lambda_3= -1.006$, and $\Lambda_4= -31.62$. There is a specific foliated splitting locus issued from the $3$-cell  $\gamma_{40}$ revealing the uncommon structure of this attractor. Since the thick S-unit and this splitting locus are local features, they do  not lead to a second positive Lyapunov exponent.  The foliated splitting locus can be identified as follows. There is a single ingoing $3$-cell which shares a $1$-cell with more than one outgoing $2$-cells: there is therefore a splitting chart without critical point as commonly observed in three-dimensional attractors, but with a critical line. 

\subsection{The Deng toroidal chaotic attractor}

\begin{figure}[ht]
  \begin{tabular}{cc}
    \includegraphics[width=0.48\linewidth]{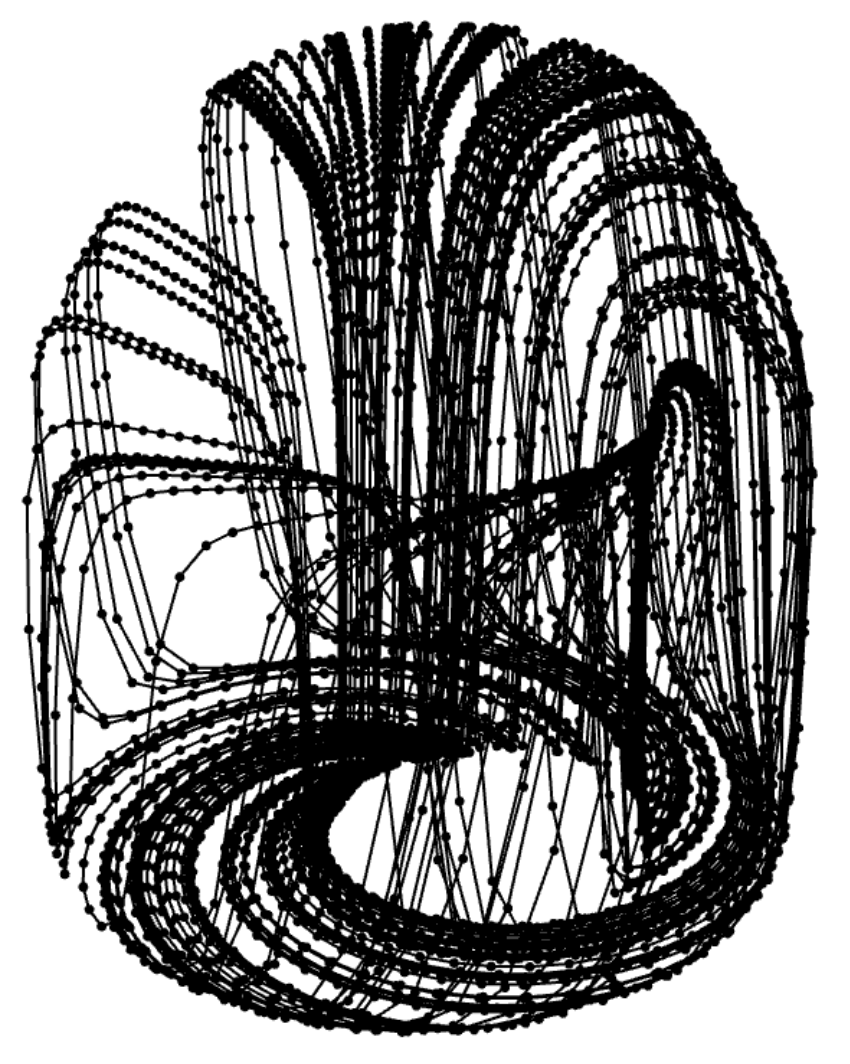} &
    \includegraphics[width=0.50\linewidth]{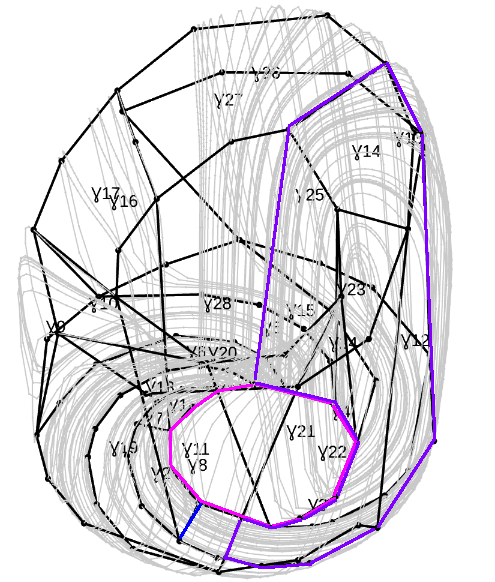} \\
    (a) Trajectory &
    (b) Original complex $K_{\rm D}$ \\[+0.2cm]
  \end{tabular}
  \includegraphics[width=0.4\linewidth]{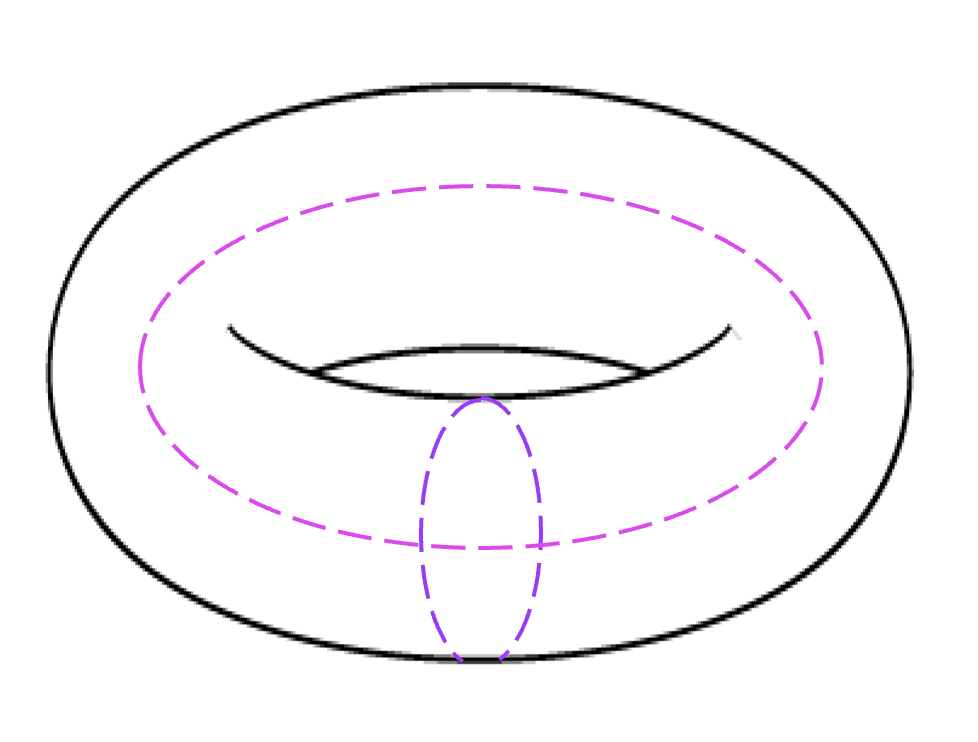} \\
	(c) Sketch of a torus T$^2$ \\[-0.2cm]
  \caption{ Chaotic toroidal attractor produced by the Deng system (\ref{CH_equation}) (a) and its cell complex $K_{\rm D}$ (b). The two $1$-generators of $H_1 [K_{\rm D}]$ ($g_1$ and $g_2$), are colored in purple and magenta, on the cell complex and on the regular torus respectively. The cell complex encloses a cavity: it has a single $2$-generator in $H_2 [K_{\rm D}]$. Parameter values: $a = 3$, $b = 0.8$, $c = 1$, $d = 0.1$, $e = 0.05$, $\eta = 3.312$, $R = 10$, $\alpha = 2.8$, $\beta = 5$, $\epsilon = 0.1$, and $\lambda = -2$. Initial conditions: $x_0 = y_0 = z_0 = 0.01$.}
  \label{CH_Deng}
\end{figure}

The Deng toroidal system\cite{Den94} is 
governed by the three ordinary differential equations 
  \begin{equation}
    \label{CH_equation}
    \left\{
      \begin{array}{l}
  	    \dot{x} =  (\lambda x - \mu y) \, z 
  	    + (2 - z) \left[ \displaystyle \alpha 
  	    \left( \displaystyle 1 - \frac{x^2 + y^2}{R^2} \right)  \, x
  	    - \beta y \right] \\[0.4cm]
  	    \dot{y} = ( \lambda y + \mu x) \, z 
  	    + (2 - z) \left[ \displaystyle \alpha 
  	    \left( \displaystyle 1 - \frac{x^2 + y^2}{R^2} \right) \, y 
  	    + \beta x \right] \\[0.3cm]
  	    \dot{z} = \displaystyle \frac{1}{\epsilon}  
  	     \left[z ((2-z) ( a (z-2)^2 + b) - dx) \right. \\[0.3cm]
  	     \hspace{1.5cm} \left. \times \left( z 
  	     + e \left( \displaystyle x^2 + y^2 \right) - \eta \right) 
  	     - \epsilon c (z - 1) \right] \, . 
      \end{array}
    \right.
  \end{equation}
For appropriate parameter values, this system produces the unimodal toroidal chaotic Deng attractor shown in Fig.\ \ref{CH_Deng}(a). 

The templex for this attractor $T_{\rm D} = (K_{\rm D}, G_{\rm D})$ is here 
proposed for the first time. Complex $K_{\rm D}$ is composed of twenty eight $2$-cells and has the homology groups of a torus, i.e. two $1$-generators $H_1 [K_{\rm D}] \sim \Z^2$ [drawn in Figs.\ \ref{CH_Deng}(b) and \ref{CH_Deng}(c)] and a single $2$-generator $H_2 [K_{\rm D}] \sim \Z$ defining the enclosed toroidal cavity. $K_{\rm D}$ presents a joining locus [Fig.\ \ref{CH_Deng}(b)] as required for chaos. 

\begin{figure}[ht]
  \includegraphics[width=\linewidth]{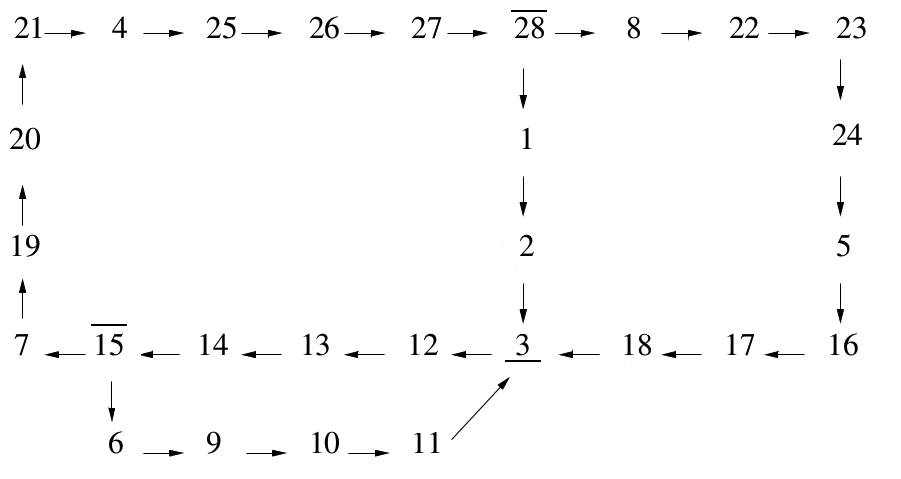} \\
	(a) Digraph $G_{\rm D}$\\[0.2cm]
	  \includegraphics[width=0.40\linewidth]{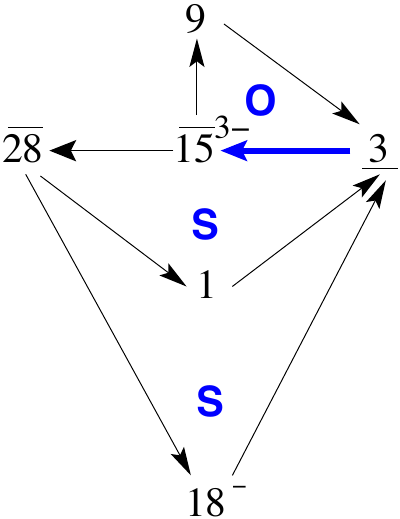} \\
	(b) Reduced digraph $\overline{G}_{\rm D}$  \\[-0.2cm]
  \caption{Digraphs for (a) the complex $K_{\rm D}$, (b) for the reduced 
complex $\overline{K}_{\rm D}$, and the three units which are identified in 
$\overline{G}_{\rm D}$. The bond is the blue arrow.}
  \label{grafoTC}
\end{figure}

Digraph $G_{\rm D}$ is shown in Fig.\ \ref{grafoTC}(a): it can be reduced to a 
six-node digraph $\overline{G}_{\rm D}$ presented in Fig.\ \ref{grafoTC}(b). 
Node $1$ is kept to prevent the digraph from having a single cell between a joining locus and a splitting locus. $\overline{K}_{\rm D}$ has a one-component joining locus associated with node $3$ in the reduced digraph $\overline{G}_{\rm D}$ as shown in Fig.\ \ref{grafoTC}(b). There are three order-1 cycles
\[
  \begin{array}{lr}
   \mathcal{C}_1(\overline{G}_{\rm D}) = \underline{3} \rightarrow \overline{15} \rightarrow 9 \rightarrow \underline{3} \\[0.1cm] 
   \mathcal{C}_2(\overline{G}_{\rm D})= \underline{3} \rightarrow \overline{15} 
<  \rightarrow \overline{28} \rightarrow 1 \rightarrow \underline{3} \\[0.1cm]
   \mathcal{C}_3(\overline{G}_{\rm D})= \underline{3} \rightarrow \overline{15} 
	  \rightarrow \overline{28} \rightarrow 18 \rightarrow \underline{3} 
  \end{array}
\]
corresponding to the three stripexes in $\overline{T}_{\rm D}$ labeled  $\mathcal{S}_1(\overline{T}_{\rm D})$, $ \mathcal{S}_2(\overline{T}_{\rm D})$ and $\mathcal{S}_3(\overline{T}_{\rm D})$. There is only one bond
\[ \mathcal{B}(\overline{G}_{\rm D}) = \{ ( \underline{3},  \overline{15} ) \} \, .
\]
$\mathcal{C}_1$ has the minimum number of nodes and is the closest to the 
rotation center, qualifying it as the only O-unit of the templex. Subtracting the edges of the O-unit from the two remaining cycles leads to the two S-units $(\overline{15}\rightarrow \overline{28} \rightarrow 1 \rightarrow  \underline{3})$ and $( \overline{28} \rightarrow 18 \rightarrow \underline{3})$ in accordance with Fig.\ \ref{grafoTC}(c). 

\begin{figure}[ht]
  \begin{tabular}{cc}
    \includegraphics[width=0.79\linewidth]{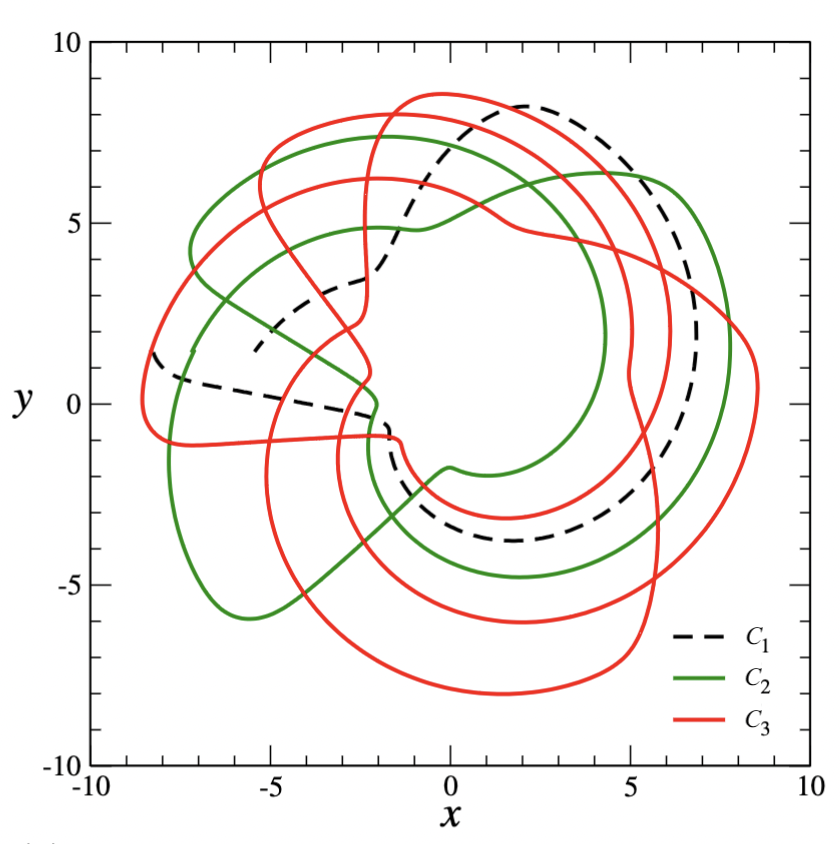} \\
	  (a) Trajectory visiting the cells of each cycle ${\cal C}_i$   \\[0.2cm]
    \includegraphics[width=0.79\linewidth]{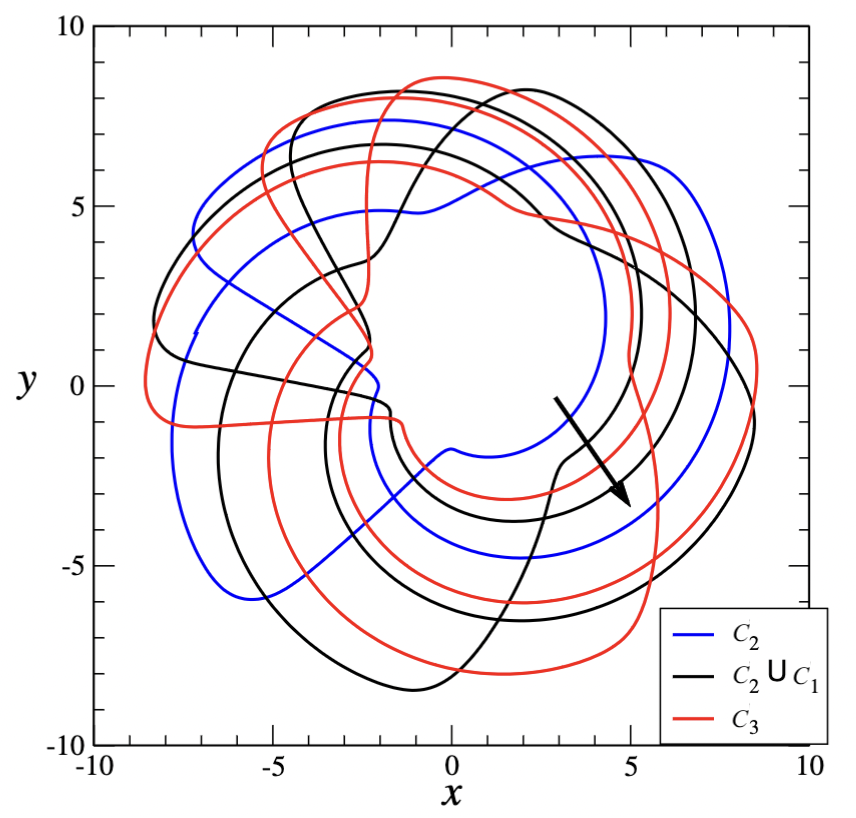} \\
    (b) Unstable periodic orbits \\[-0.2cm]
  \end{tabular}
  \caption{Unstable periodic orbits extracted from the Deng toroidal chaotic 
attractor (a) and the parts of the trajectory associated with each cycle (b). 
The joining locus is drawn as a thick arrow in (b).}
  \label{dengupos}
\end{figure}

The Deng attractor is structured around a torus T$^2$ and, consequently, it is governed by two main frequencies as confirmed by a Fourier spectrum leading to $f_1 = 1.022$~Hz and $f_2 = 2.046$~Hz, with a ratio $\frac{f_1}{f_2} = 0.4997 \approx 0.5$. Each of these frequencies is associated with one of the two $1$-generators of the homology group $H_1(\overline{K}_{\rm D})$, that structure a torus as sketched in Fig.\ \ref{CH_Deng}(c).
To better understand how stripexes $\mathcal{S}_1(\overline{T}_{\rm D})$, $ \mathcal{S}_2(\overline{T}_{\rm D})$ and $\mathcal{S}_3(\overline{T}_{\rm D})$ are organized, we extract three excerpts of chaotic trajectory visiting each of the three stripexes. The duration associated with these excerpts is $\tau_1 = 0.926$~s, $\tau_2 = 1,980$~s, and  $\tau_3 = 2.916$~s, respectively [see Fig.\ \ref{dengupos}(a)].

\begin{figure}[ht]
  \includegraphics[width=0.22\textwidth]{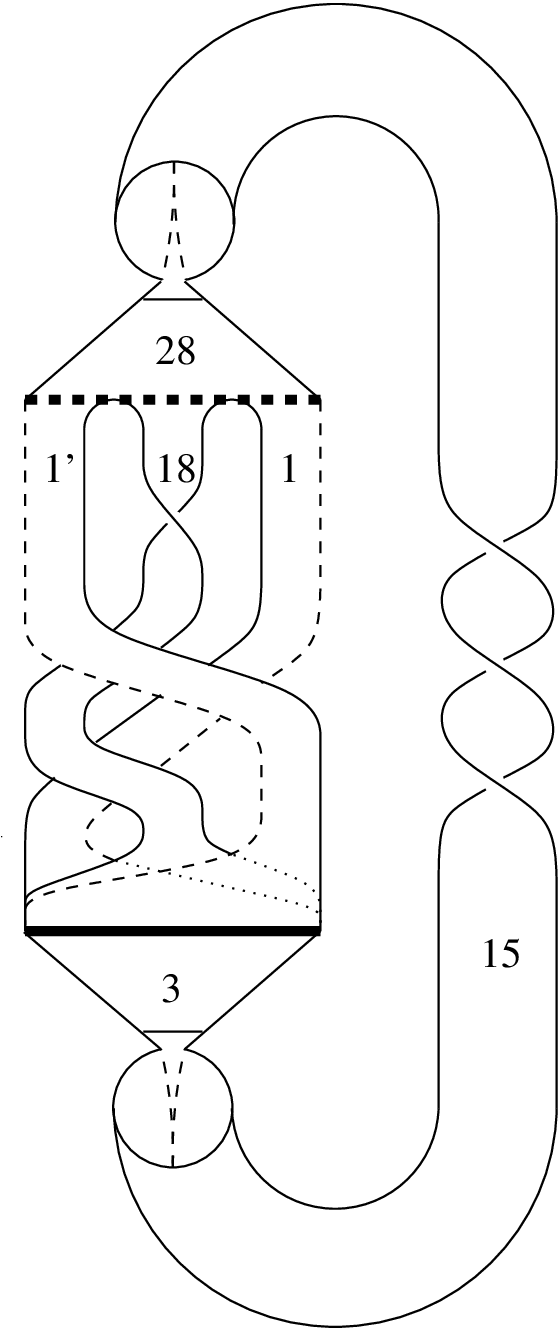} \\[-0.2cm]
  \caption{Direct template for the toroidal Chaotic Deng attractor with some of the $2$-cells appearing in the reduced digraph [Fig.\ \ref{grafoTC}(b)]. (Redrawn from Fig. 14 in Ref. \onlinecite{Man21a}). The splitting locus is dashed, and the joining locus is thick.}
  \label{dengtemtoko}
\end{figure}

By using a close-return technique applied to a common Poincaré section of the attractor, we find one period-$2$ and two period-$3$ orbits [Fig.\ \ref{dengupos}(b)]. From $\overline{G}_{\rm D}$, they correspond to cycles ${\cal C}_2$, ${\cal C}_2 \cup {\cal C}_1$, and ${\cal C}_3$, respectively. Cycles ${\cal C}_2$, and ${\cal C}_3$ are of order-$1$ and they should be, by definition, associated with period-1 orbits.  Since cycle ${\cal C}_1$ corresponds to the main frequency associated with a revolution around the main hole of the toroidal structure ($1$-generator $g_1$), the corresponding period-1 orbit should be associated with the $1$-generator $g_1$, that is, with a trajectory in the center of the closed cavity defined by the torus: consequently, it cannot belong to the population of unstable periodic orbits embedded within the attractor. This means that the common Poincaré section is not the optimal one for investigating this toroidal chaotic flow. Using a Poincaré section based on the joining locus, the period-$3$ orbit corresponding to ${\cal C}_2\cup {\cal C}_1$ is now a period-$2$ orbit as expected for an orbit visiting two stripexes. By reducing the Poincaré section to the joining locus, the orbit periods are now equal to the number of times they visit a stripex.  If we use the three symbols $\{ 1, 2, 3 \}$ to designate the three cycles ${\cal C}_i$, the extracted orbits are encoded by $(2)$, $(21)$, and $(3)$, respectively. The period-$1$ orbit $(1)$ is missing, as explained above. This result is a strong confirmation that a unimodal toroidal chaos is associated with three stripexes and is hence described by the three-strip template in Fig.\ \ref{dengtemtoko} as proposed in Ref.\ \onlinecite{Man21a}.

\section{Conclusion and discussion}
\label{conc}

A templex is a flow-based {\sc BraMAH} complex approximating the structure on which the attractor lies, endowed with a digraph indicating how the flow visits the locally highest dimensional cells. Reducing the templex by merging unnecessary cells and eliminating the corresponding nodes allows to define classes of chaotic attractors at the penultimate taxonomic rank, that is, at the level of the first-return map. By considering various types of chaotic dynamics ---  unimodal or multimodal, with symmetry or not, three-dimensional or higher, with toroidal underlying structure or not ---, we here show that the reduced templex can be interpreted in terms of two elementary dynamical processes, namely, oscillating units (O-units) and switching units (S-units). The way these two types of units are combined can be directly extracted from the reduced templex. Our procedure has the advantage of not being limited to three-dimensional state space and not requiring the definition of a Poincaré section. As a result, splitting and joining loci are accurately identified. The toroidal chaos case exhibits the generality of these two loci types proving that the joining locus 
should be used to defined the Poincaré section.

The number and nature of the units are derived from the properties of the stripex set, which are manipulated as cycles from the reduced digraph. When the dynamics can be synthesized by a template, the units can be recognized directly using the cell labels on the strips.

We thus show that it is indeed possible to decompose chaotic systems into elementary dynamical processes using a templex and a reduction procedure to algorithmically obtain the bonds between O-units and S-units. This is a significant advance for a complete taxonomy of chaos. Note that the reduced digraph does not contain all the information for an ultimate discrimination between topologically non-equivalent attractors, but the templex does. The reduced digraph can be dressed with additional information in reporting the local twists of the stripexes, as well as additional (extrinsic) properties such as the number of half-twists. The ultimate taxonomic rank is provided by the templex properties (or by the template in the dissipative three-dimensional limit).  The characterization of other types of toroidal chaos with these theoretical tools is currently under study. The case of hyperchaos is also one of our next tasks. A paper devoted to the templex extraction from experimental data is currently underway to show that this method is not limited to theoretical cases. For particularly complex attractors where trajectories may remain trapped for a long time interval, the templex may vary depending on the time series being analyzed. In such cases, it would be beneficial to explore different domains of the attractor from well-selected time series to generate a templex for each part of the attractor, which can then be combined to provide a comprehensive description of the whole attractor.

\section*{Acknowledgments}
This work has received funding from the ANR project TeMPlex ANR-23-CE56-0002 (D.S.). G.D.C. gratefully acknowledges her postdoctoral scholarship within the ANR project. C.M. gratefully acknowledges her doctoral scholarship from CONICET. C. L. and coworkers thank N. Stankevich for the Lyapunov exponents computation.

\bibliography{SysDyn}

\appendix

\section{Cell complexes and homology groups}
\label{algebraictop}

In algebraic topology, a $k$-cell within a cell complex $K$ is a set that can be mapped continuously and invertibly to the interior of a $k$-disk, with its boundary divided into a finite number of lower-dimensional cells, called faces. As a whole, the cell complex $K$ is a finite collection of cells where the faces of each cell are also elements of $K$, and the interiors of distinct cells do not overlap. The dimension $\kappa$ of a cell complex $K$ is defined by the dimension of its highest-dimensional cells. A cell complex is termed directed or oriented if each cell is assigned a direction. In a directed complex, a $k$-chain is a linear combination of $k$-cells, where the coefficients are integers. These chains capture the connectivity of the cells at each $k$ level.

\begin{figure} 
  \includegraphics[width=0.9\linewidth]{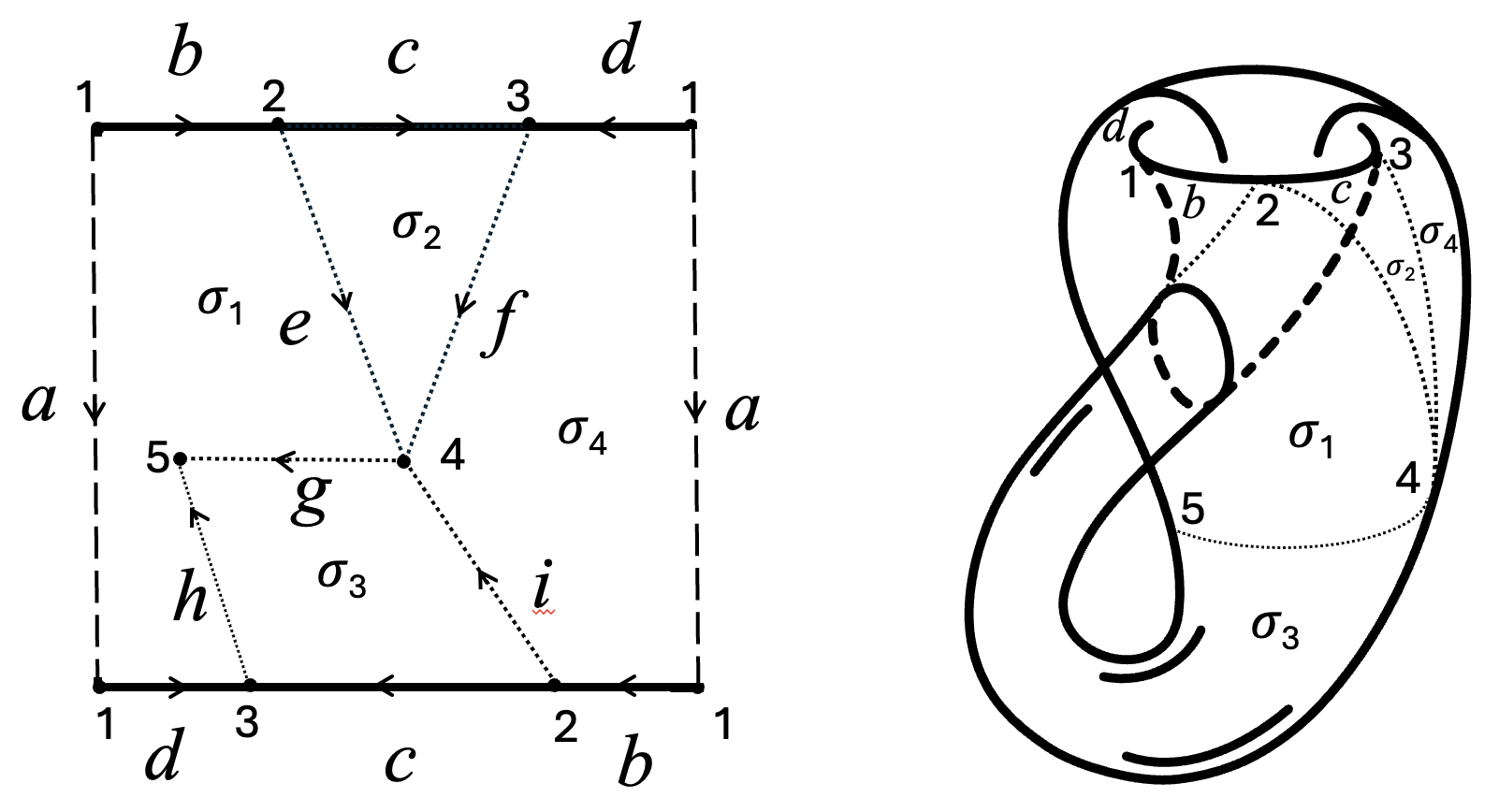} \\
  \caption{A cell complex $K$ with four $2$-cells $\sigma_i$ ($i=1 \dots 4$) for a Klein bottle. The repeated labels in the planar diagram indicate gluing instructions to form the Klein bottle in four dimensions. }
  \label{Klein}
\end{figure}

\begin{figure} 
  \includegraphics[width=0.9\linewidth]{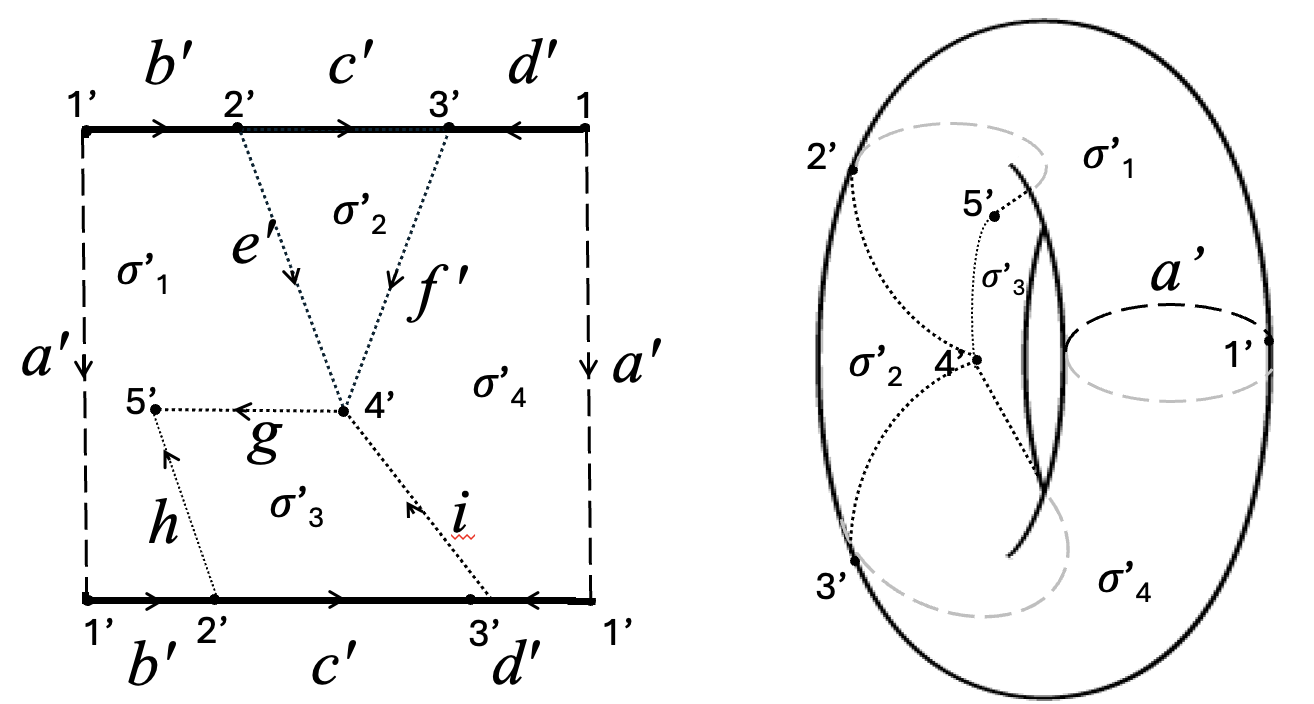} \\
  \caption{A cell complex $K'$ with four $2$-cells $\sigma'_i$ ($i=1 \dots 4$) for a torus. The repeated labels in the planar diagram indicate gluing instructions to form the toroidal surface in three dimensions. }
  \label{Torus}
\end{figure}

Let us consider the example of the Klein bottle in Figure \ref{Klein}.  A cell complex is provided, with four $2$-cells $\sigma_i$ ($i=1 \dots 4$), nine $1$-cells ($a$, $b$, $c$, $d$, $e$, $f$, $g$, $h$, $i$) and five $0$-cells ($\langle 1\rangle$, $\langle 2\rangle$, $\langle 3\rangle$, $\langle 4\rangle$, $\langle 5 \rangle$). Repeated labels are important in a planar diagram since they indicate gluing instructions. Notice that with the same set of cells and labels but changing some of their positions and directions, one can get a torus, as shown in Figure \ref{Torus}. 

\begin{defiNN}
Two $k$-chains are homologous if their difference is the border of some $(k + 1)$-chain.
\end{defiNN}

\noindent
In our example, $c$, $f$ and $e$ are such that $c+f$ and $e$ are homologous ($c+f \sim e$), since $c + f - e = \partial(\sigma_2)$. The $2$-cell $\sigma_2$ serves as a bridge between the edges $c+f$ and $e$, implying that any information conveyed by $c+f$ is also represented by $e$, and the reverse holds true as well. The homology equivalence relation leads to the definition of homology groups.

\begin{defiNN}
For a directed complex $K$, the $k$-dimensional homology group of $K$ is the quotient group $H_k(K) = Z_k(K)/B_k(K)$, i.e. the group of equivalence classes of elements of $Z_k(K)$ with the homology relation.
\end{defiNN}

\noindent
The homologically independent $k$-dimensional cycles form the $k$-generators of $H_k$, also called $k$-holes. The $k$th Betti number, denoted as $\beta_k$, is the rank of $H_k$, i.e. the number of elements in the basis. In simple terms, $\beta_0$ indicates the number of connected components in the complex, $\beta_1$ reflects the number of closed paths that encircle holes, and the number of enclosed cavities is captured by $\beta_2$. At each level, generators can be used to identify each of the connected components, holes or cavities.

For the cell complex of the Klein bottle, there is a single $0$-generator or connected component and so $\beta_0(K)=1$, $H_0(K)  \sim \Z$ with basis $\{ \langle 1 \rangle \}$. At level $1$, we have $H_1(K)  \sim \Z$ with basis  $\{ a \}$, and $\beta_1(K)=1$. Finally, at level $2$, there are no enclosed cavities i.e. $H_2(K)  \sim 0$ and $\beta_2(K)=0$. In fact, since $(b+c-d)$ must be followed around twice to form a cycle, the surface is said to have a torsion coefficient of $2$.

The torus has Betti numbers $\beta_0(K')=1$, $\beta_1(K')=2$ and $\beta_2(K')=1$. More precisely, $H_0(K')  \sim \Z$ with basis $\{\langle 1 \rangle\}$, $H_1(K') = \sim \Z^2$ with basis $\{a', b'+c'-d'\} $ (there are two $1$-holes), and $H_2(K') \sim \Z$ with basis $\{ \sigma'_1+\sigma'_2+\sigma'_3+\sigma'_4 \}$, enclosing the cavity. The torus has no torsion coefficients.

\begin{figure}
  \includegraphics[width=0.58\linewidth]{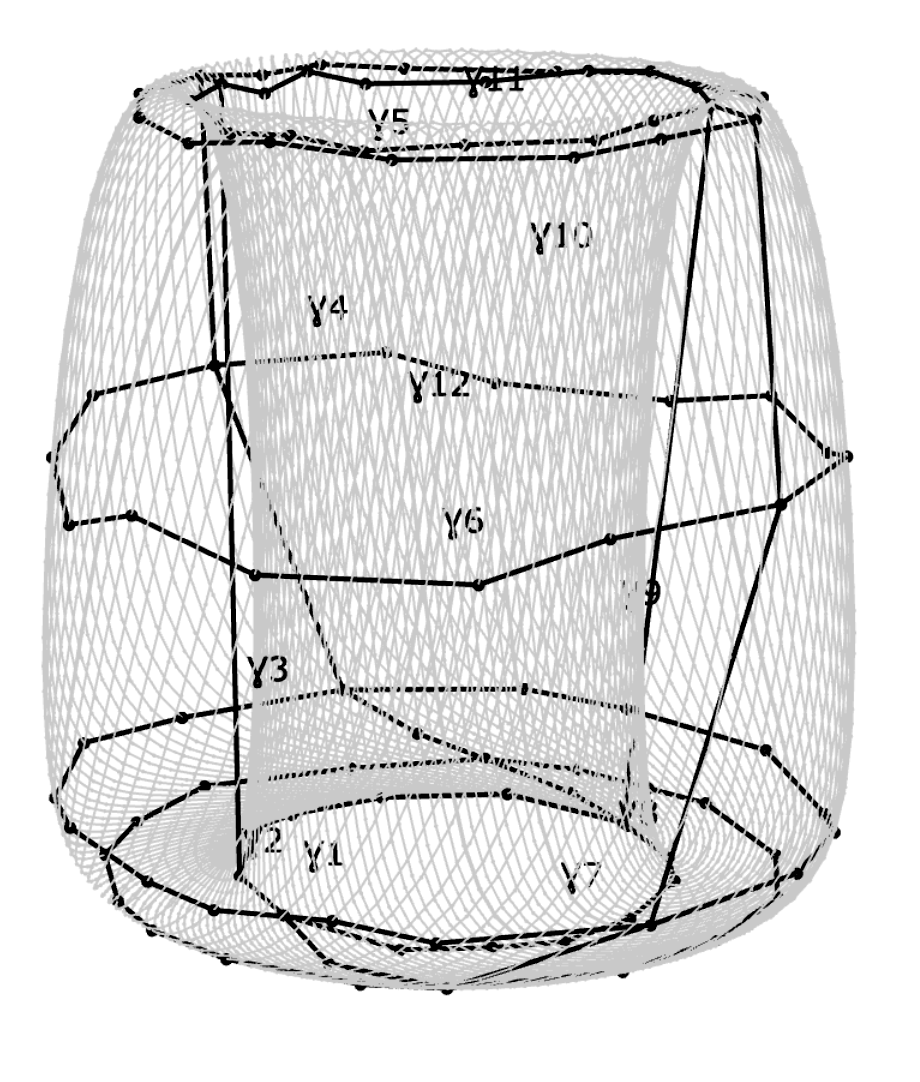} \\
	(a) Original complex $K_{\rm QP}$ \\[0.2cm]
  \includegraphics[width=0.60\linewidth]{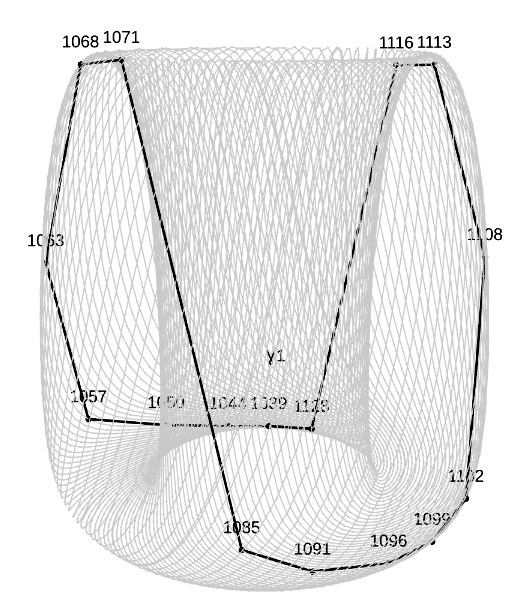} \\
	(b) Reduced complex $\overline{K}_{\rm QP}$ \\[-0.2cm]
  \caption{Original $K_{\rm QP}$ and reduced $\overline{K}_{\rm QP}$ complexes 
for the quasi-periodic attractor produced by the system (\ref{QP_equation}). 
The reduced complex has a single $2$-cell labeled $\gamma_1$.}
  \label{QP_Deng}
\end{figure}

\begin{figure} 
  \includegraphics[width=0.11\linewidth]{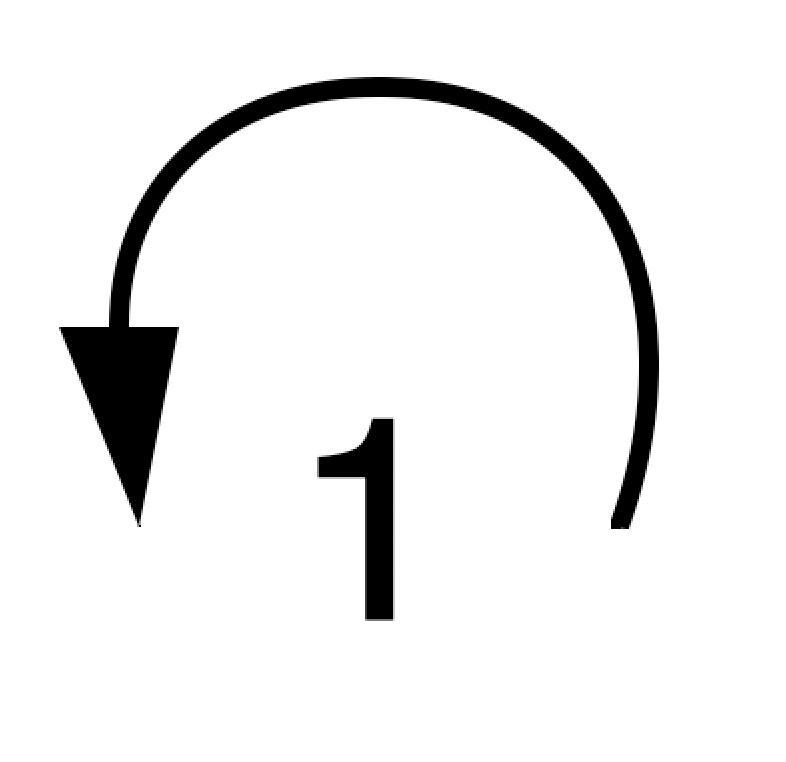} \\
	Reduced digraph $\overline{G}_{\rm QP}$
  \caption{Digraph $\overline{G}_{\rm QP}$ for the quasi-periodic attractor produced by the system (\ref{QP_equation}) for the single $2$-cell labeled $\gamma_1$.}
  \label{Self}
\end{figure}

\section{Templex for regular dynamics}
\label{regdyn}

A torus is well represented by complex $K'$ in appendix \ref{algebraictop} but can also be analyzed with a complex formed by a single $2$-cell with at least two $1$-cells, as shown in figures $4.13$ and $4.14$ in \citet{Kin93}, page $62$. The fact that a single $2$-cell is sufficient to build up the complex of a torus highlights the fact that a cell complex is not a simple mesh or  tessellation of an object, and that each layered dimension $k$ contains important information. When the regular surface of a torus is traveled by a quasi-periodic trajectory characterized by a templex without joining locus, the relevant information is already encoded in the cell complex, since there are no distinct ways for flowing within its structure.

When the original templex has no joining locus, the underlying flow is periodic  or quasi-periodic, and the manifold supporting the trajectories has no `branches'. In other words, there are no alternative paths for the flow along  the underlying manifold, or equivalently, there are no stripexes. In this case, the templex can be simplified by reducing the number of highest dimensional cells in the complex (and therefore the nodes in the digraph) as long as homologies and torsion groups do not change. The highest dimensional cells of the original complex can be merged till we get a complex with the minimal number of $\kappa$-cells leading to the same homology and torsion groups. Thus, a single $2$-cell is sufficient to describe a torus. And a regular flow on a torus can be represented by a reduced digraph with a self-edge representing the unique path of the flow. 

An example of such a flow is provided by the Deng toroidal system\cite{Den94}  formed by the following three ordinary differential equations:
\begin{equation}
  \label{QP_equation}
  \left\{
    \begin{array}{l}
      \dot{x} = \displaystyle z (\lambda x - \mu y) 
	    + (2 - z) \left[ \displaystyle \alpha x 
	    \left( \displaystyle 1 - \frac{x^2 + y^2}{R^2} \right) - \beta y
	    \right] \\[0.4cm]
	    \dot{y} = \displaystyle z ( \lambda y + \mu x) + (2 - z) 
	    \left[ \displaystyle \alpha y 
	      \left( \displaystyle 1 - \frac{x^2 + y^2}{R^2} \right) 
	      + \beta x \right] \\[0.5cm]
	    \dot{z} =  \displaystyle z (2-z)  
	    \frac{ \left[ \displaystyle z + e \left( x^2 + y^2 \right) 
	    - \eta \right] - \epsilon c (z - 1)}{\epsilon}  
    \end{array}
  \right.
\end{equation}

The symmetry of the system (\ref{QP_equation}) prevents it from developing chaos-prone folds,\cite{Den94} so that the resulting system is quasi-periodic (Fig. \ref{QP_Deng}). As expected, complex $K_{\rm QP}$ has no joining locus. Its homology groups are those of a torus: $K_{\rm QP}$ has two $1$-generators, that is, $H_1 [K_{\rm QP}] \sim \Z^2$ and a $2$-generator for its enclosed cavity $H_2 [K_{\rm QP}] \sim \Z$. The $2$-cells covering the torus surface can all be merged into a single $2$-cell, as shown in Fig. \ref{QP_Deng}(b), without altering the homology groups. There are no torsion groups. The reduced digraph is a single node (corresponding to $\gamma_1$) with a single edge from it to itself, displaying the trivial flow organization on the torus. The digraph is shown in Fig.~\ref{Self}.  

\end{document}